\documentclass[12pt]{article}
\usepackage{makeidx}
\usepackage{amssymb}
\usepackage{amsfonts}
\usepackage{amsmath}
\usepackage{graphicx}
\usepackage{setspace}
\usepackage{authblk}
\usepackage{units}
\usepackage{appendix}
\usepackage[left=2cm,top=2.3cm,right=2cm,bottom=2.1cm]{geometry}
\usepackage{xcolor}
\usepackage[hyperfootnotes=false]{hyperref}

\renewcommand\familydefault\rmdefault

\hypersetup{colorlinks=true,linkcolor=black,citecolor=black,urlcolor=black}

\begin{document}

\title{{\textbf{Singular electromagnetic fields in nonlinear electrodynamics with a constant background field}}}
\author[1]{T. C. Adorno\thanks{Tiago.Adorno@xjtlu.edu.cn}}
\author[2,3]{D. M. Gitman\thanks{gitman@if.usp.br}}
\author[2]{A. E. Shabad\thanks{shabad@lpi.ru}}
\affil[1]{\textit{Department of Physics, Xi'an Jiaotong-Liverpool University, 111 Ren'ai Road, Suzhou Dushu Lake Science and Education Innovation District, Suzhou Industrial Park, Suzhou, 215123, P. R. China;}}
\affil[2]{\textit{P. N. Lebedev Physical Institute, 53 Leninskiy prospekt, 119991, Moscow, Russia;}}
\affil[3]{\textit{Instituto de F\'{\i}sica, Universidade de S\~{a}o Paulo, Caixa Postal 66318, CEP 05508-090, S\~{a}o Paulo, S.P., Brazil.}}

\maketitle

\onehalfspacing

\begin{abstract}
When exploring equations of nonlinear electrodynamics in effective medium
formed by mutually parallel external electric and magnetic fields, we come
to special static axial-symmetric solutions of two types. The first are
comprised of fields referred to as electric and magnetic responses to a
point-like electric charge when placed into the medium. In
electric case, this is a field determined by the induced charge density.
In magnetic case, this is a field carrying no magnetic charge and determined by an induced current. Fields of second type
require presence of pseudoscalar constants for their existence. These are
singular on the axis drawn along the external fields. In electric case this
is a field of an inhomogeneously charged infinitely thin thread. In magnetic
case this is the magnetic monopole with the Dirac string supported by
solenoidal current. In both cases the necessary pseudoscalar constant is
supplied by field derivatives of nonlinear Lagrangian taken on external
fields. There is also a magnetic thread solution dual to electric thread
with null total magnetic charge.
\end{abstract}


\section{Introduction\label{Sec1}}

The search for nonlinear effects stimulated by fields is a subject of
widespread interest across diverse areas of physics. In Quantum
Electrodynamics (QED), for instance, the nonlinearity supplied by strong
electromagnetic fields -- which has been known ever since the
seminal works by Klein \cite{Klein}, Sauter \cite{Sauter}, Heisenberg and
Euler \cite{Heisenberg}, Weiskpoff \cite{Weisskopf}, Euler and K\"{o}ckel 
\cite{Kockel}, and Schwinger \cite{Schwinger51} -- is attracting
considerable attention lately owing to advances in laser technology, driven
especially by chirped pulse amplification \cite{StrMou85}. Thanks to this,
there is a growing number of ongoing and planned experiments devoted to
detecting nonlinear quantum phenomena stemming from ultra-intense focused
laser pulse collisions. The subject is well discussed in a number of reviews 
\cite%
{MarShu06,Dunne09,Piazza12,BatRiz13,KinHei16,Blackburn20,Karbstein20,GonBlaMarBul22,Fedotov-et-all-23}
and references therein (see also the references \cite%
{Grib,Greiner,FGS,Dittrich,Gies,Dunne-rev,Ruffini,GelTan} for theoretical
foundations of QED with external fields).

From a fundamental perspective, the nonlinearity in QED is provided \ by the
scattering of light by light, which is described by the four-photon diagram.
Its experimental observation in collisions of nuclei was recently reported
in \cite{light}. On the other hand, it is long since understood that\
nonlinear process of photon splitting into two \cite{Adler}, \cite{merging}
shows itself in formation of radiation \cite{thompson} produced by highly
magnetized neutron stars (pulsars). Owing to the nonlinearity, the vacuum
filled with strong electromagnetic fields\ makes an equivalent medium
described by polarization and mass operators calculated through Feynman
diagrams with \textquotedblleft dressed\textquotedblright\ propagators for
electrons and positrons, i.e., in the Furry representation \cite{Furry},
with exact solutions of the Dirac equation in external-field background\ for
them. The common primary goal for observation \cite{birefringence} of
nonlinearity of QED is the birefringence of vacuum \cite{Adler}, \cite{Erber}%
, \cite{BatSha71} filled with a strong field (notably, the recent evidence
for it is obtained from a neutron star \cite{birefringence2})).

While the rigorous treatment of nonlinear quantum effects in QED requires
the knowledge of polarization tensors calculated within the Furry
representation, essential simplification can be achieved by computing them
with the aid of effective Lagrangians in the local approximation, depending
exclusively on field strengths (and not their space-time derivatives). This
approximation, which supplements Maxwell's electrodynamics with effective
actions, provides an excellent framework for investigating phenomenological
consequences of the nonlinear character of the vacuum with external fields.

Models based on nonlinear electrodynamics are gaining considerable attention
in diverse areas, such as in Einstein's General Relativity (GR) coupled to
electromagnetic fields. In this instance, the influence on the Universe's
evolution is under study, and a variety of new types of charged black holes
appear, among which the magnetized black holes are closest to the authors'
interests; see, e.g., \cite{kruglov} and references therein\footnote{%
See also Ref. \cite{Kim23} for recent studies on vacuum birefringence
effects near magnetars.}. In the context of particle physics, there is an
intense activity regarding the experimental search for magnetic
monopoles basing on experiments on high-energy peripheral nucleus-nucleus
collisions, wherein pure photon-photon interaction is distinguishable, whose
consideration resulted in discovery of the long-waited-for light-by-light
scattering \cite{ATLAS}. The same photon-photon interaction may become
responsible for monopole-antimonopole pairs production. Among nonlinear
models recruited to treat the photon-photon interaction, Born-Infeld
extensions of the Standard Model attract special attention (see e.g. \cite{EllMavYou17,MavMit22,EllMavRolTev22}), as following from string theory \cite{Tseytlin} and providing finiteness of the field energy to a point-like
monopole, both electric or magnetic, identified with its mass\footnote{There are also other nonlinear models that guarantee convergence \cite{Costa et al} of the field energies of monopoles (even if their fields are
singular), the Euler-Heisenberg Lagrangian taken to quadratic order of its
expansion in powers of the field invariants belonging to this class.
Sufficient conditions to be imposed on the growth of nonlinearity with the
field may be found in \cite{Shishmarev et al}.}. See also Refs. \cite%
{GinSch99,EpeFanCanMitVen12,ColTer16} for alternative models and Refs. \cite%
{MavMit17,Mitsou22,Mitsou22b,Mitsou23} for experimental searches within
MoEDAL's collaboration \cite{moedal16,moedal18,moedal19}. There is also the
possibility of monopole-antimonopole pair creation from the vacuum by strong
magnetic fields produced during {color{red}{the mentioned}} peripheral
collisions. \cite{GouRaj17,Rajantie19,GouHoRaj19,HoRaj20,GouHoRaj21}.
Besides these subjects, models based on nonlinear electrodynamics were also
considered in radiation problems e.g. \cite{GaeHel23}, systems coupled to
axion fields/dark-matter models e.g. \cite%
{Snowmass13,ChaShaMu21,ForSin22,Helayel22}, and even in condensed matter
physics \cite{Helayel23}.

In the present paper, we study nonlinear Maxwell's equations for
electromagnetic fields produced by a point charge placed in
strong electromagnetic fields\footnote{%
Other manifestations of the magneto-electric effect were treated in \cite%
{ShaGit2012}, \cite{AdoGitSha2013} \cite{AdoGitSha2014} as nonlinear
response.}. We are basing on the general local nonlinear electrodynamics,
whose action is an arbitrary nonlinear functional of electric and magnetic
field strengths independent of their time- and space-derivatives. Examples
of such theories are provided by the so-called Heisenberg-Euler action
calculated within one-loop \cite{Heisenberg} and two-loop \cite{twoloop}
accuracy in QED, by the famous Born-Infeld \cite{born}\ action and by many
other models considered together with GR, see references in \cite{kruglov}.
All results are expressed in terms of derivatives over fields of the general
nonlinear local Lagrangian.

In Sec. \ref{Sec2} perturbation expansion to solutions of nonlinear Maxwell
equations with respect of small nonlinearity is outlined, and the electric\
charge and current density induced by the point-like electric charge are
obtained against the external-field background. We find, first, the electric
response to that charge, which reduces to its screening and to inducing an
angle-distributed electric charge density. Next, we find an angle-depending
magnetic response due to the induced current. It carries no magnetic charge {%
and it is depicted in Fig. \ref{Fig2} (see also \cite{AGS2020}%
, \cite{Preprint}). In Sec. \ref{Sec3}, we supplement the responses of Sec. %
\ref{Sec2} with arbitrary-magnitude solutions of corresponding homogeneous
equations. These combined fields suffer singularities on lines parallel to
the common direction of the background fields. The electric contribution to
the combined field is found in Subsection\ref{Sec3.1}. It requires to be
supported by a charge concentrated on an infinitely thin thread, and an
outer pseudoscalar constant formed by the background fields is needed. The
magnetic contribution to the combined field is found in Subsection\ref{Ss3.2}
This is a magnetic monopole rigged with the field contained in an infinitely
thin cylinder -- a version of the Dirac string. This solution requires to be
supported by magnetic charge concentrated in a point and by a solenoidal
current encircling that cylinder. It also requires a pseudoscalar constant
-- as this is common with magnetic monopoles. There is yet another magnetic
solution dual to the electric thread of Subsection \ref{Sec3.1}. In this
instance, no pseudoscalar constant is needed, since this solution is
intrinsically a pseudovector, as a magnetic field should be.

We work in the four-dimensional Minkowski space-time with coordinates $%
X=\left( X^{\mu },\ \mu =0,i\right) =\left( t,\mathbf{r}\right) $, $t=X^{0}$%
, $r=X^{i}=\left( x,y,z\right) $, $i=1,2,3$, and metric tensor $\eta _{\mu
\nu }=\mathrm{diag}\left( +1,-1,-1,-1\right) $.\ The fully antisymmetric
four- and three-dimensional Levi-Civita tensors are normalized as $%
\varepsilon ^{0123}=+1$\ and $\varepsilon ^{123}=+1$, respectively. Natural
units ($\hslash =1=c$) are used.

\section{Induced 4-currents and electromagnetic responses to point electrict
charge in constant background fields\label{Sec2}}

\subsection{Maxwell equations and perturbations}

We consider nonlinear electrodynamics given by a local Lagrangian $L\left(
X\right) =-\mathfrak{F}\left( X\right) +\mathfrak{L}\left( X\right) $, i.e.,
the one depending on 4-coordinates $X$\ via relativistic-invariant
combinations of electromagnetic field strengths $\mathfrak{F}=F_{\mu \nu
}F^{\mu \nu }/4=\left( \mathbf{B}^{2}-\mathbf{E}^{2}\right) /2$ and $%
\mathfrak{G}=F_{\mu \nu }\widetilde{F}^{\mu \nu }/4=-\mathbf{B}\cdot \mathbf{%
E}$, but independent of their space- or time-derivatives\footnote{%
The most common example of such theory is provided by the famous
Euler-Heisenberg Lagrangian taken for $\mathfrak{L}\left( X\right) $, which
is the approximation of the effective Lagrangian of Quantum Electrodynamics
fit for slow-varying fields (it is known in literature as calculated with
the accuracy of one and two electron-positron loops).}. Nonlinear Maxwell's
equations are obtained using the least action principle in the form:%
\begin{equation}
\partial ^{\nu }\left[ \left( 1-\frac{\delta \mathfrak{L}\left( \mathfrak{F},%
\mathfrak{G}\right) }{\delta \mathfrak{F}\left( X\right) }\right) F_{\nu \mu
}\left( X\right) -\frac{\delta \mathfrak{L}\left( \mathfrak{F},\mathfrak{G}%
\right) }{\delta \mathfrak{G}\left( X\right) }\tilde{F}_{\nu \mu }\left(
X\right) \right] =j_{\mu }\left( X\right) \,,  \label{MaxEq}
\end{equation}%
where the dual field is defined as $\widetilde{F}^{\mu \nu }=\varepsilon
^{\mu \nu \alpha \beta }F_{\alpha \beta }/2$, and the current $j_{\mu
}\left( X\right) $ is introduced into the action in the standard way%
\begin{equation}
S\left[ A\right] =\int \left[ L\left( X\right) -j_{\mu }\left( X\right)
A^{\mu }\left( X\right) \right] dX\,,\ \ dX=dtd\mathbf{r}\,,\ \ d\mathbf{r}%
=dxdydz\,,  \label{L}
\end{equation}%
with $A^{\mu }\left( X\right) $ being the potential for electromagnetic
field, $F^{\mu \nu }\left( X\right) =\partial ^{\mu }A^{\nu }\left( X\right)
-\partial ^{\nu }A^{\mu }\left( X\right) $.

We shall consider electromagnetic fields against a background of constant
and homogeneous external field $\overline{F}^{\mu \nu }$. Small
electromagnetic deviations {$f^{\mu \nu }\left( X\right) =F^{\mu \nu }\left(
X\right) -\overline{F}^{\mu \nu }$} from the background are described by
linearized Maxwell's equations obtained from (\ref{MaxEq})\footnote{%
Derivation of Eq. (\ref{eq1}) from the action (\ref{L}) is described in
detail in Refs. \cite{ShaGit2012,AdoGitSha2013,AGS2020,AdoGitSha2016}.}:%
\begin{eqnarray}
\partial ^{\nu }f_{\nu \mu }\left( X\right) -j_{\mu }\left( X\right)
&=&\partial ^{\tau }\left[ \mathfrak{L}_{\mathfrak{F}}f_{\tau \mu }\left(
X\right) +\frac{1}{2}\left( \mathfrak{L}_{\mathfrak{FF}}\overline{F}_{\alpha
\beta }+\mathfrak{L}_{\mathfrak{FG}}\widetilde{\overline{F}}_{\alpha \beta
}\right) \overline{F}_{\tau \mu }f^{\alpha \beta }\left( X\right) \right] 
\notag \\
&+&\partial ^{\tau }\left[ \mathfrak{L}_{\mathfrak{G}}\tilde{f}_{\tau \mu
}\left( X\right) +\frac{1}{2}\left( \mathfrak{L}_{\mathfrak{FG}}\overline{F}%
_{\alpha \beta }+\mathfrak{L}_{\mathfrak{GG}}\widetilde{\overline{F}}%
_{\alpha \beta }\right) \widetilde{\overline{F}}_{\tau \mu }f^{\alpha \beta
}\left( X\right) \right] \,.  \label{eq1}
\end{eqnarray}

In (\ref{eq1}),%
\begin{eqnarray}
&&\mathfrak{L}_{\mathfrak{F}}=\frac{\partial \overline{\mathfrak{L}}}{%
\partial \overline{\mathfrak{F}}}\,,\ \ \mathfrak{L}_{\mathfrak{G}}=\frac{%
\partial \overline{\mathfrak{L}}}{\partial \overline{\mathfrak{G}}}\,,\ \ {%
\mathfrak{L}_{\mathfrak{FF}}=\frac{\partial ^{2}\overline{\mathfrak{L}}}{%
\partial \overline{\mathfrak{F}}^{2}}}\,,  \notag \\
&&{\mathfrak{L}_{\mathfrak{GG}}}={\frac{\partial ^{2}\overline{\mathfrak{L}}%
}{\partial \overline{\mathfrak{G}}^{2}}}\,,\ \ \mathfrak{L}_{\mathfrak{FG}}=%
\frac{\partial ^{2}\overline{\mathfrak{L}}}{\partial \overline{\mathfrak{F}}%
\partial \overline{\mathfrak{G}}}\,,  \label{eq2}
\end{eqnarray}%
are ($X$-independent)\ derivatives of the effective Lagrangian $\overline{%
\mathfrak{L}}=\mathfrak{L}\left( \overline{\mathfrak{F}},\overline{\mathfrak{%
G}}\right) $, taken on the constant background field invariants $\mathfrak{F}%
=\overline{\mathfrak{F}}=\overline{F}_{\mu \nu }\overline{F}^{\mu \nu
}/4=\left( \overline{\mathbf{B}}^{2}-\overline{\mathbf{E}}^{2}\right) /2$
and $\mathfrak{G}=\overline{\mathfrak{G}}=\overline{F}^{\mu \nu }\widetilde{%
\overline{F}}_{\mu \nu }/4=-\overline{\mathbf{B}}\cdot \overline{\mathbf{E}}$%
\textrm{. }When comparing (\ref{MaxEq}) with (\ref{eq1}) it should be noted
that the variational derivatives have given way to ordinary ones --
according, for example, to the rule\textrm{\ }$\left. \frac{\delta \mathfrak{%
L}\left( \mathfrak{F},\mathfrak{G}\right) }{\delta \mathfrak{F}\left(
X\right) }\right\vert _{f^{\nu \mu }=0}=\frac{\partial \overline{\mathfrak{L}%
}\left( \overline{\mathfrak{F}},\overline{\mathfrak{G}}\right) }{\partial 
\overline{\mathfrak{F}}}$ \textit{etc.} -- when calculating the first term
of Taylor expansion in powers of $f^{\nu \mu }$ in the process of
linearization of equation (\ref{MaxEq})\textrm{.}

The set of nonlinear field equations (\ref{MaxEq}) and their linearization (%
\ref{eq1})\ are manifestly gauge invariant, as they involve electromagnetic
field strength tensors only, and not potentials. These sets are both in
concord with gauge invariance, because the latter demands the
4-transversality of the current $\partial ^{\mu }j_{\mu }\left( X\right) =0$%
, assumed to be fulfilled. Applying partial derivative operator $\partial
^{\mu }$ to the left-hand sides of (\ref{MaxEq}), or (\ref{eq1}), yields
zero due the antisymmetricity of the electromagnetic field strength tensors $%
F_{\alpha \beta }=-F_{\beta \alpha }$, $f^{\alpha \beta }=-f^{\beta \alpha }$%
.

Lagrangian $L\left( X\right) $ is relativistic-invariant, while Eq. (\ref%
{MaxEq}) is relativistic-covariant due to the fact that $L\left( X\right) =-%
\mathfrak{F}(X)+\mathfrak{L}\left( \mathfrak{F}(X),\mathfrak{G}(X)\right) $ depends
on relativistic-invariant combinations of fields$.$ On the contrary, Eq. (%
\ref{eq1}) is not relativistic-invariant, since it describes evolution
against the background of external field $\overline{F}_{\mu \nu }$. As long
as$\ \overline{\mathfrak{G}}\neq 0$, there exists the reference frame, where 
$\overline{\mathbf{B}}\parallel \pm \overline{\mathbf{E}}$. Hence, solutions
of Eq. (\ref{eq1}) retain invariance under rotation around common direction
of the external electric and magnetic fields.

It is important to note that the current $j_{\mu }\left( X\right) $ in (\ref%
{eq1}) is the same, as it was in (\ref{MaxEq}). This is because the
left-hand side (\ref{MaxEq}) disappears on the constant field $F_{\mu \nu }=%
\overline{F}_{\mu \nu }$ (zeroth term, $f^{\nu \mu }=0$, in the Taylor
expansion). In other words, the constant external field requires no current,
while\ the current $j_{\mu }\left( X\right) $ is a source of small
perturbation to the background. For this reason electromagnetic deviations $%
f_{\nu \mu }\left( X\right) $\ are also termed \textquotedblleft linear
response functions,\textquotedblright\ when they refer to how the effective
\textquotedblleft medium\textquotedblright\ formed by external fields\
reacts to small imposed current $j^{\mu }\left( X\right) $. Simultineously,
the right-hand side in Eq. (\ref{eq1}) should be referred as current,
\textquotedblleft induced\textquotedblright\ in the \textquotedblleft
medium\textquotedblright\ by the small perturbation $j_{\mu }\left( X\right) 
$.

Henceforth, we treat Eq. (\ref{eq1}) by perturbations relative to the
nonlinearity. This is meaningful for weakly nonlinear theories with their
nonlinear part of Lagrangian small\footnote{%
as in QED, whose effective Lagrangians are proportional to the fine
structure constant $\alpha =e^{2}/4\pi $.} as compared to linear one, $%
\left\vert \mathfrak{L}\right\vert /\left\vert \mathfrak{F}\right\vert \ll 1$%
. To build the perturbative series, we formally multiply the right-hand side
of Eq. (\ref{eq1}) by a parameter $\epsilon $ that represents this
smallness, and seek solutions for (\ref{eq1}) in power-series of $\epsilon $:%
\begin{equation}
f_{\nu \mu }\left( X\right) =\sum_{n=0}^{\infty }\epsilon ^{n}f_{\nu \mu
}^{\left( n\right) }\left( X\right) \,.  \label{eq3}
\end{equation}%
The zeroth-order terms $f_{\nu \mu }^{\left( 0\right) }\left( X\right) $ are
Maxwell's fields, solutions to equations%
\begin{equation}
\partial ^{\nu }f_{\nu \mu }^{\left( 0\right) }\left( X\right) =j_{\mu
}\left( X\right) \,,  \label{eq2b}
\end{equation}%
representing the inhomogeneous set of standard Maxwell's equations for the
deviations. Being independent of nonlinearity, $f_{\nu \mu }^{\left(
0\right) }\left( X\right) $ dominates over other terms in expansion (\ref%
{eq3}). The next-to-leading terms obey the recurrence relations%
\begin{eqnarray}
&&\partial ^{\nu }\left[ f_{\nu \mu }^{\left( n+1\right) }\left( X\right) -%
\mathfrak{L}_{\mathfrak{F}}f_{\nu \mu }^{\left( n\right) }\left( X\right) -%
\mathfrak{L}_{\mathfrak{G}}\tilde{f}_{\nu \mu }^{\left( n\right) }\left(
X\right) -\mathfrak{D}^{\left( n\right) }\left( X\right) \overline{F}_{\nu
\mu }-\mathfrak{S}^{\left( n\right) }\left( X\right) \widetilde{\overline{F}}%
_{\nu \mu }\right] =0\,,\ \ n\geq 0\,,  \notag \\
&&\mathfrak{D}^{\left( n\right) }\left( X\right) =\frac{\mathfrak{L}_{%
\mathfrak{FF}}\overline{F}^{\alpha \beta }+\mathfrak{L}_{\mathfrak{FG}}%
\widetilde{\overline{F}}^{\alpha \beta }}{2}f_{\alpha \beta }^{\left(
n\right) }\left( X\right) \,,\ \ \mathfrak{S}^{\left( n\right) }\left(
X\right) =\frac{\mathfrak{L}_{\mathfrak{FG}}\overline{F}_{\alpha \beta }+%
\mathfrak{L}_{\mathfrak{GG}}\widetilde{\overline{F}}_{\alpha \beta }}{2}%
f_{\alpha \beta }^{\left( n\right) }\left( X\right) \,.  \label{eq4}
\end{eqnarray}%
All extra integration constants to arise when treating the first-order
differential equations (\ref{eq4}) should be set equal to zero.

Restricting ourselves to stationary charge distributions $j_{\mu }\left(
X\right) =j_{\mu }\left( \mathbf{r}\right) $ and electromagnetic fields $%
f_{\nu \mu }\left( X\right) =f_{\nu \mu }\left( \mathbf{r}\right) $,
Maxwell's equations for linear responses (\ref{eq4}) have the form%
\begin{eqnarray}
&\boldsymbol{\nabla }\cdot \left[ \mathbf{E}^{\left( n+1\right) }\left( 
\mathbf{r}\right) -\boldsymbol{\mathfrak{E}}^{\left( n\right) }\left( 
\mathbf{r}\right) \right] =&0\,,  \notag \\
&\boldsymbol{\nabla }\times \left[ \mathbf{B}^{\left( n+1\right) }\left( 
\mathbf{r}\right) -\boldsymbol{\mathfrak{H}}^{\left( n\right) }\left( 
\mathbf{r}\right) \right] =&0\,,  \label{eq4.1}
\end{eqnarray}%
where $\boldsymbol{\mathfrak{E}}^{\left( n\right) }\left( \mathbf{r}\right) $
and $\boldsymbol{\mathfrak{H}}^{\left( n\right) }\left( \mathbf{r}\right) $
are auxiliary fields,%
\begin{eqnarray}
\boldsymbol{\mathfrak{E}}^{\left( n\right) }\left( \mathbf{r}\right) &=&%
\mathfrak{L}_{\mathfrak{F}}\mathbf{E}^{\left( n\right) }\left( \mathbf{r}%
\right) +\mathfrak{L}_{\mathfrak{G}}\mathbf{B}^{\left( n\right) }\left( 
\mathbf{r}\right) +\mathfrak{D}^{\left( n\right) }\left( \mathbf{r}\right) 
\overline{\mathbf{E}}+\mathfrak{S}^{\left( n\right) }\left( \mathbf{r}%
\right) \overline{\mathbf{B}}\,,  \label{eq4.2} \\
\boldsymbol{\mathfrak{H}}^{\left( n\right) }\left( \mathbf{r}\right) &=&%
\mathfrak{L}_{\mathfrak{F}}\mathbf{B}^{\left( n\right) }\left( \mathbf{r}%
\right) -\mathfrak{L}_{\mathfrak{G}}\mathbf{E}^{\left( n\right) }\left( 
\mathbf{r}\right) +\mathfrak{D}^{\left( n\right) }\left( \mathbf{r}\right) 
\overline{\mathbf{B}}-\mathfrak{S}^{\left( n\right) }\left( \mathbf{r}%
\right) \overline{\mathbf{E}}\,.  \label{eq4.3}
\end{eqnarray}%
that supply the Maxwell equations (\ref{eq4.1}) with linearly-induced
current densities\newline
$j_{\mu }^{\left( n+1\right) }\left( \mathbf{r}\right) =\left( \rho ^{\left(
n+1\right) }\left( \mathbf{r}\right) ,\mathbf{j}^{\left( n+1\right) }\left( 
\mathbf{r}\right) \right) $,%
\begin{equation}
\boldsymbol{\nabla }\cdot \boldsymbol{\mathfrak{E}}^{\left( n\right) }\left( 
\mathbf{r}\right) =\rho ^{\left( n+1\right) }\left( \mathbf{r}\right) \,,\ \ 
\boldsymbol{\nabla }\times \boldsymbol{\mathfrak{H}}^{\left( n\right)
}\left( \mathbf{r}\right) =\mathbf{j}^{\left( n+1\right) }\left( \mathbf{r}%
\right) \,.  \label{eq4.6}
\end{equation}

For parallel background fields $\overline{\mathbf{B}}\parallel \overline{%
\mathbf{E}}$, say, aligned with the unit vector $\mathbf{\hat{z}}$, ($%
\left\vert \mathbf{\hat{z}}\right\vert =1$), the auxiliary fields admit the
forms%
\begin{eqnarray}
\boldsymbol{\mathfrak{E}}^{\left( n\right) }\left( \mathbf{r}\right) &=&%
\mathfrak{L}_{\mathfrak{F}}\mathbf{E}^{\left( n\right) }\left( \mathbf{r}%
\right) +\mathfrak{L}_{\mathfrak{G}}\mathbf{B}^{\left( n\right) }\left( 
\mathbf{r}\right) +\mathfrak{b}E_{\parallel }^{\left( n\right) }\left( 
\mathbf{r}\right) \mathbf{\hat{z}}+\mathfrak{\tilde{g}}B_{\parallel
}^{\left( n\right) }\left( \mathbf{r}\right) \mathbf{\hat{z}\,,}
\label{eq4.4} \\
\boldsymbol{\mathfrak{H}}^{\left( n\right) }\left( \mathbf{r}\right) &=&%
\mathfrak{L}_{\mathfrak{F}}\mathbf{B}^{\left( n\right) }\left( \mathbf{r}%
\right) -\mathfrak{L}_{\mathfrak{G}}\mathbf{E}^{\left( n\right) }\left( 
\mathbf{r}\right) +\mathfrak{c}B_{\parallel }^{\left( n\right) }\left( 
\mathbf{r}\right) \mathbf{\hat{z}}-\mathfrak{\tilde{g}}E_{\parallel
}^{\left( n\right) }\left( \mathbf{r}\right) \mathbf{\hat{z}}\,,
\label{eq4.5}
\end{eqnarray}%
where $E_{\parallel }^{\left( n\right) }\left( \mathbf{r}\right) =\mathbf{%
\hat{z}}\cdot \mathbf{E}^{\left( n\right) }\left( \mathbf{r}\right) $, $%
B_{\parallel }^{\left( n\right) }\left( \mathbf{r}\right) =\mathbf{\hat{z}}%
\cdot \mathbf{B}^{\left( n\right) }\left( \mathbf{r}\right) $, and $%
\mathfrak{b}$, $\mathfrak{c}$, $\mathfrak{\tilde{g}}$ are %
dimensionless combinations of the Lagrangian derivatives
and the field invariants,%
\begin{eqnarray}
\mathfrak{b} &=&-\mathfrak{L}_{\mathfrak{FF}}\overline{\mathbf{E}}^{2}-%
\mathfrak{L}_{\mathfrak{GG}}\overline{\mathbf{B}}^{2}+2\overline{\mathfrak{G}%
}\mathfrak{L}_{\mathfrak{FG}}\,,  \notag \\
\mathfrak{c} &=&\mathfrak{L}_{\mathfrak{FF}}\overline{\mathbf{B}}^{2}+%
\mathfrak{L}_{\mathfrak{GG}}\overline{\mathbf{E}}^{2}+2\overline{\mathfrak{G}%
}\mathfrak{L}_{\mathfrak{FG}}\,,  \notag \\
\mathfrak{\tilde{g}} &=&\left( \mathfrak{L}_{\mathfrak{GG}}-\mathfrak{L}_{%
\mathfrak{FF}}\right) \overline{\mathfrak{G}}+2\overline{\mathfrak{F}}%
\mathfrak{L}_{\mathfrak{FG}}\,.  \label{gtilded}
\end{eqnarray}%
Here $\mathfrak{b}$, $\mathfrak{c}$ are scalars, and $\mathfrak{\tilde{g}}$
is a pseudoscalar. Under the dual transformation $\overline{\mathbf{E}}%
\rightarrow i\overline{\mathbf{B}},$ $\overline{\mathbf{B}}\rightarrow -i%
\overline{\mathbf{E}},$ the pseudoscalar $\mathfrak{\tilde{g}}$\ is
invariant, while $\mathfrak{b}$ and $\mathfrak{c}$ map into one another: $%
\mathfrak{b}\leftrightarrow \mathfrak{c}.$%

Along with Eqs. (\ref{eq4.1}), the electromagnetic responses of
all orders adhere to the Bianchi identities:%
\begin{equation}
\partial _{\mu }\tilde{f}^{\left( n\right) \mu \nu }\left( \mathbf{r}\right)
=0\rightarrow \boldsymbol{\nabla }\times \mathbf{E}^{\left( n\right) }\left( 
\mathbf{r}\right) =\mathbf{0}\,,\ \ \boldsymbol{\nabla }\cdot \mathbf{B}%
^{\left( n\right) }\left( \mathbf{r}\right) =0\,.  \label{eq8}
\end{equation}%
These identities stem from the formulation of the theory in terms of potentials.

\subsection{Responses to point electric charge}

In what follows we focus on studying electromagnetic responses to a
pointlike charged particle $j_{\mu }\left( x\right) =q\delta _{\mu 0}\delta
\left( \mathbf{r}\right) $, $\delta \left( \mathbf{r}\right) =\delta \left(
x\right) \delta \left( y\right) \delta \left( z\right) $ placed in the
background field. This distribution produces a Coulomb field%
\begin{equation}
f_{0i}^{\left( 0\right) }\left( \mathbf{r}\right) =E^{\left( 0\right)
i}\left( \mathbf{r}\right) =\frac{qx^{i}}{4\pi r^{3}}\,,\ \ B^{\left(
0\right) i}\left( \mathbf{r}\right) =-\frac{1}{2}\varepsilon
^{ijk}f_{jk}^{\left( 0\right) }=0\,,  \label{eq5}
\end{equation}%
which is a solution of Eqs. (\ref{eq2b}). According to Eqs. (\ref{eq4.1}),
first-order electromagnetic responses $\mathbf{E}^{\left( 1\right) }\left( 
\mathbf{r}\right) $ and $\mathbf{B}^{\left( 1\right) }\left( \mathbf{r}%
\right) $ obey the set of equations%
\begin{eqnarray}
&&\boldsymbol{\nabla }\cdot \left[ \mathbf{E}^{\left( 1\right) }\left( 
\mathbf{r}\right) -\boldsymbol{\mathfrak{E}}^{(0)}\left( \mathbf{r}\right) %
\right] =0\,,  \label{eq6.1} \\
&&\boldsymbol{\nabla }\times \left[ \mathbf{B}^{\left( 1\right) }\left( 
\mathbf{r}\right) -\boldsymbol{\mathfrak{H}}^{(0)}\left( \mathbf{r}\right) %
\right] =0\,,  \label{nnew7b}
\end{eqnarray}%
whose auxiliary fields $\boldsymbol{\mathfrak{E}}^{(0)}\left( \mathbf{r}%
\right) $, $\boldsymbol{\mathfrak{H}}^{(0)}\left( \mathbf{r}\right) $, for
parallel background fields $\mathbf{\hat{z}}\parallel \overline{\mathbf{B}}%
\parallel \overline{\mathbf{E}}$, read%
\begin{equation}
\boldsymbol{\mathfrak{E}}^{(0)}\left( \mathbf{r}\right) =\frac{q}{4\pi r^{2}}%
\left( \mathfrak{L}_{\mathfrak{F}}\mathbf{\hat{r}}+\mathfrak{b}\cos \theta 
\mathbf{\hat{z}}\right) \,,\ \ \boldsymbol{\mathfrak{H}}^{(0)}\left( \mathbf{%
r}\right) =-\frac{q}{4\pi r^{2}}\left( \mathfrak{L}_{\mathfrak{G}}\mathbf{%
\hat{r}}+\mathfrak{\tilde{g}}\cos \theta \mathbf{\hat{z}}\right) \,.
\label{eq7}
\end{equation}%
Here, $\theta $ denotes the angle between the unit radius vector $\mathbf{%
\hat{r}}=\mathbf{r}/\left\vert \mathbf{r}\right\vert $ and the direction of
the external fields, $\theta =\arccos \left( \mathbf{\hat{z}}\cdot \mathbf{%
\hat{r}}\right) $. Eq. (\ref{eq7}) only holds for the case of mutually
parallel background fields. It is obtained directly from (\ref{eq4.4}, \ref%
{eq4.5}), which in their turn, follow from (\ref{eq4.2}, \ref{eq4.3}) with
account of (\ref{eq4}). 
Eq. (\ref{eq7}) only holds for the case of mutually parallel background fields
. It is obtained directly from (\ref{eq4.4}), (\ref{eq4.5})
, which in their turn, follow from (\ref{eq4.2}), (\ref{eq4.3}) with account of (\ref{eq4}).

Previously \cite{AGS2020,AdoGitSha2016}, we found that the first-order
electromagnetic responses are projections of the auxiliary fields (\ref{eq7}%
) -- which guarantees fulfilment of Bianchi identities for them at least in
nonsingular points --%
\begin{eqnarray}
&&\mathbf{E}^{\left( 1\right) }\left( \mathbf{r}\right) =\boldsymbol{\nabla }%
\frac{1}{\boldsymbol{\nabla }^{2}}\left[ \boldsymbol{\nabla }\cdot 
\boldsymbol{\mathfrak{E}}^{\left( 0\right) }\left( \mathbf{r}\right) \right]
=\frac{q}{4\pi r^{2}}\left[ \left( \mathfrak{L}_{\mathfrak{F}}+\frac{1-3\cos
^{2}\theta }{2}\mathfrak{b}\right) \mathbf{\hat{r}}+\mathfrak{b}\cos \theta 
\mathbf{\hat{z}}\right] \,,  \label{eq9.1} \\
&&\mathbf{B}^{\left( 1\right) }\left( \mathbf{r}\right) =\boldsymbol{%
\mathfrak{H}}^{(0)}\left( \mathbf{r}\right) -\boldsymbol{\nabla }\frac{1}{%
\boldsymbol{\nabla }^{2}}\left[ \boldsymbol{\nabla }\cdot \boldsymbol{%
\mathfrak{H}}^{(0)}\left( \mathbf{r}\right) \right] =\frac{q\mathfrak{\tilde{%
g}}}{8\pi }\left( \frac{1-3\cos ^{2}\theta }{r^{2}}\right) \mathbf{\hat{r}}%
\,.  \label{ansa20}
\end{eqnarray}

\subsubsection{Electric response}

We observed that the electric field (\ref{eq9.1}) is \textit{charged}%
\footnote{%
This is only for brevity. To be precise, we had to say \textquotedblleft
electric field has the induced charge as its source\textquotedblright . We shall take
the liberty to apply such abuse of terminology to magnetic fields as well.},
as the corresponding induced charge density $\rho ^{\left( 1\right) }\left( 
\mathbf{r}\right) $ is%
\begin{equation}
\rho ^{\left( 1\right) }\left( \mathbf{r}\right) =\boldsymbol{\nabla }\cdot 
\mathbf{E}^{\left( 1\right) }\left( \mathbf{r}\right) =q\left( \mathfrak{L}_{%
\mathfrak{F}}+\frac{\mathfrak{b}}{3}\right) \delta \left( \mathbf{r}\right) +%
\frac{q\mathfrak{b}}{4\pi r^{3}}\left( 1-3\cos ^{2}\theta \right) \,.
\label{eq10}
\end{equation}%
The delta-function contribution, obtained in \cite{AdoGitSha2016}, is
determined by the auxiliary field $\boldsymbol{\mathfrak{E}}^{\left(
0\right) }\left( \mathbf{r}\right) $ through Gauss' theorem%
\begin{equation}
\int_{V}\boldsymbol{\nabla }\cdot \boldsymbol{\mathfrak{E}}^{\left( 0\right)
}\left( \mathbf{r}\right) d\mathbf{r}=\oint_{S}\boldsymbol{\mathfrak{E}}%
^{\left( 0\right) }\left( \mathbf{r}\right) \cdot \mathbf{\hat{n}}dS\,,
\label{eq11.1}
\end{equation}%
evaluated for an sphere $S$, with volume $V$, centered at the point charge $%
q $. The induced total charge $q^{\left( 1\right) }$ is nontrivial%
\begin{equation}
q^{\left( 1\right) }=\int_{V}\rho ^{\left( 1\right) }\left( \mathbf{r}%
\right) d\mathbf{r}=q\left( \mathfrak{L}_{\mathfrak{F}}+\frac{\mathfrak{b}}{3%
}\right) \,,  \label{eq11}
\end{equation}%
and concentrated at the origin $\mathbf{r}=\mathbf{0}$, due to the
delta-function part. The above identity, combined with the regular part of
Eq. (\ref{eq10}), complies with Eq. (\ref{eq6.1}), i.e., $\boldsymbol{\nabla 
}\cdot \boldsymbol{\mathfrak{E}}^{\left( 0\right) }\left( \mathbf{r}\right) =%
\boldsymbol{\nabla }\cdot \mathbf{E}^{\left( 1\right) }\left( \mathbf{r}%
\right) =\rho ^{\left( 1\right) }\left( \mathbf{r}\right) $. The charge $%
q^{\left( 1\right) }$\ is the vacuum correction to the seeded point Coulomb
charge $q$, while the distributed part of the induced charge density $q%
\mathfrak{b}\left( 1-3\cos ^{2}\theta \right) /4\pi r^{3}$\ does not
contribute to total charge inside a sphere of any radius. The induced charge
density (\ref{eq10}) is illustrated in Fig. 1 of Ref. \cite{AdoGitSha2016}.

The Bianchi identity%
\begin{equation}
\boldsymbol{\nabla }\times \mathbf{E}^{\left( 1\right) }\left( \mathbf{r}%
\right) =\mathbf{0}\,,  \label{eq11.2}
\end{equation}%
is explicitly fulfilled everywhere, except, admittedly, at $\mathbf{r}=%
\mathbf{0}$. Any circle going around the\ singular point $\mathbf{r}=\mathbf{%
0}$\ may be -- without creating any fluxes of $\boldsymbol{\nabla }\times 
\mathbf{E}^{\left( 1\right) }\left( \mathbf{r}\right) $ --$\ $continuously
transformed to one situated in the plane, orthogonal to the $z$-axis\ and
containing the point $\mathbf{r}=\mathbf{0}$.\ The resulting integral $%
\oint_{\mathcal{C}}\mathbf{E}^{\left( 1\right) }\left( \mathbf{r}\right)
\cdot d\mathbf{l}$ evidently disappears owing to the explicit form (\ref%
{eq9.1}), since $\mathbf{\hat{r}}\cdot d\mathbf{l}=\mathbf{\hat{z}}\cdot d%
\mathbf{l}=0$ for such a circle.\ Once radius of the circle can be made
arbitrarily small, we see that no flux of $\boldsymbol{\nabla }\times 
\mathbf{E}^{\left( 1\right) }\left( \mathbf{r}\right) $\ flows through the
point $\mathbf{r}=\mathbf{0}$.\ This implies that $\boldsymbol{\nabla }%
\times \mathbf{E}^{\left( 1\right) }\left( \mathbf{r}\right) =\mathbf{0}$\
also in this point. Hence, the electric response (\ref{eq9.1}) may be
represented everywhere as the gradient of a scalar potential, $\mathbf{E}%
^{\left( 1\right) }\left( \mathbf{r}\right) =-\boldsymbol{\nabla }%
A_{0}^{\left( 1\right) }\left( \mathbf{r}\right) $,%
\begin{equation}
A_{0}^{\left( 1\right) }\left( \mathbf{r}\right) =\frac{q}{4\pi r}\left( 
\mathfrak{L}_{\mathfrak{F}}+\frac{1-\cos ^{2}\theta }{2}\mathfrak{b}\right)
\,.  \label{eq11b}
\end{equation}

\subsubsection{Magnetic response}

As for the magnetic response (\ref{ansa20}), it is supplied by a nontrivial
induced\ current density%
\begin{equation}
\mathbf{j}^{\left( 1\right) }\left( \mathbf{r}\right) =\boldsymbol{\nabla }%
\times \mathbf{B}^{\left( 1\right) }\left( \mathbf{r}\right) =\boldsymbol{%
\nabla }\times \boldsymbol{\mathfrak{H}}^{\left( 0\right) }\left( \mathbf{r}%
\right) =\frac{3q}{4\pi r^{3}}\mathfrak{\tilde{g}}\left( \mathbf{\hat{z}}%
\cdot \mathbf{\hat{r}}\right) \left[ \mathbf{\hat{r}}\times \mathbf{\hat{z}}%
\right] \,.  \label{par}
\end{equation}%
The current flux (\ref{par}) flows in opposite directions in the upper and
lower hemispheres, as illustrated in Fig. 1 of Ref. \cite{AGS2020}. Hence,
the current flux accross the part $\mathcal{S}$\ of a fixed meridian plane $%
\varphi =\varphi _{0}$, $0\leq \varphi _{0}<2\pi $,\ which is the ring
enclosed between any two circles $r_{1}<r<r_{2}$\ in that plane, is zero%
\begin{equation}
\int_{\mathcal{S}}\mathbf{j}^{\left( 1\right) }\left( \mathbf{r}\right)
\cdot \mathbf{\hat{n}}d\mathcal{S}=\frac{3q}{4\pi }\mathfrak{\tilde{g}}%
\int_{r_{1}}^{r_{2}}\frac{dr}{r^{2}}\int_{0}^{\pi }d\theta \cos \theta \sin
\theta =0\,,  \label{circ}
\end{equation}%
where $\mathbf{\hat{n}}$ denotes the unit vector normal to each point of the
chosen surface.\ So is the contour integral $\oint_{\mathcal{C}}\mathbf{B}%
^{\left( 1\right) }\left( \mathbf{r}\right) \cdot d\mathbf{l}${\Huge \ }%
along contour $\mathcal{C}$\ embracing $\mathcal{S}$\ due to Stokes'
theorem. The magnetic response (\ref{ansa20}) also obeys Bianchi's identity%
\begin{equation}
\boldsymbol{\nabla }\cdot \mathbf{B}^{\left( 1\right) }\left( \mathbf{r}%
\right) =0\,,  \label{bianchi}
\end{equation}%
extended to the point $\mathbf{r}=\mathbf{0}$ as well, that excludes an
overall magnetic charge and makes the\ formulation of the theory in terms of
potentials admissible. The fulfilment of the Bianchi identity (\ref{bianchi}%
), formally guaranteed by the projection operator in (\ref{ansa20}), can
also be -- beyond the singularity point $\mathbf{r}=0$ -- directly verified
by substitution of (\ref{ansa20}). The first-order linear magnetic response (%
\ref{ansa20}) does not carry any magnetic charge, in virtue of the
triviality of the Gauss integral%
\begin{equation}
\oint_{S}\mathbf{B}^{\left( 1\right) }\left( \mathbf{r}\right) \cdot \mathbf{%
\hat{n}}dS=0\,,  \label{arb230}
\end{equation}%
seen explicitly for a sphere $S$, centered at the charge $q$, and therefore
also valid for arbitrary surface in view of the Bianchi identity for $%
\mathbf{r}\neq \mathbf{0}$. This implies that the corresponding induced
magnetic charge density $\rho _{M}^{\left( 1\right) }\left( \mathbf{r}%
\right) $ is identically zero%
\begin{equation}
\rho _{M}^{\left( 1\right) }\left( \mathbf{r}\right) =\boldsymbol{\nabla }%
\cdot \mathbf{B}^{\left( 1\right) }\left( \mathbf{r}\right) =0\,,
\label{eq12}
\end{equation}%
everywhere, including the origin $\mathbf{r}=0$, where the formal
calculation of the divergence involved in it makes no sense. As a result,
one concludes that there is no magnetic charge attributed to the magnetic
response $\mathbf{B}^{\left( 1\right) }\left( \mathbf{r}\right) $: the
magnetic lines of force incoming to and outgoing from the charge $q$
compensate each other, so that the corresponding magnetic flux be zero. The
magnetic lines of force are straight lines, vanishing at the angles $\theta
=\pm \arccos 1/\sqrt{3}$. As no net magnetic charge exists for producing a
nontrivial magnetic flux (\ref{arb230}), there are inward magnetic lines
(pointing to $q$) and outward magnetic lines (pointing out of $q$), in the
same proportion; see Fig \ref{Fig2}.

\begin{figure}[th]
\begin{center}
\includegraphics[scale=0.38]{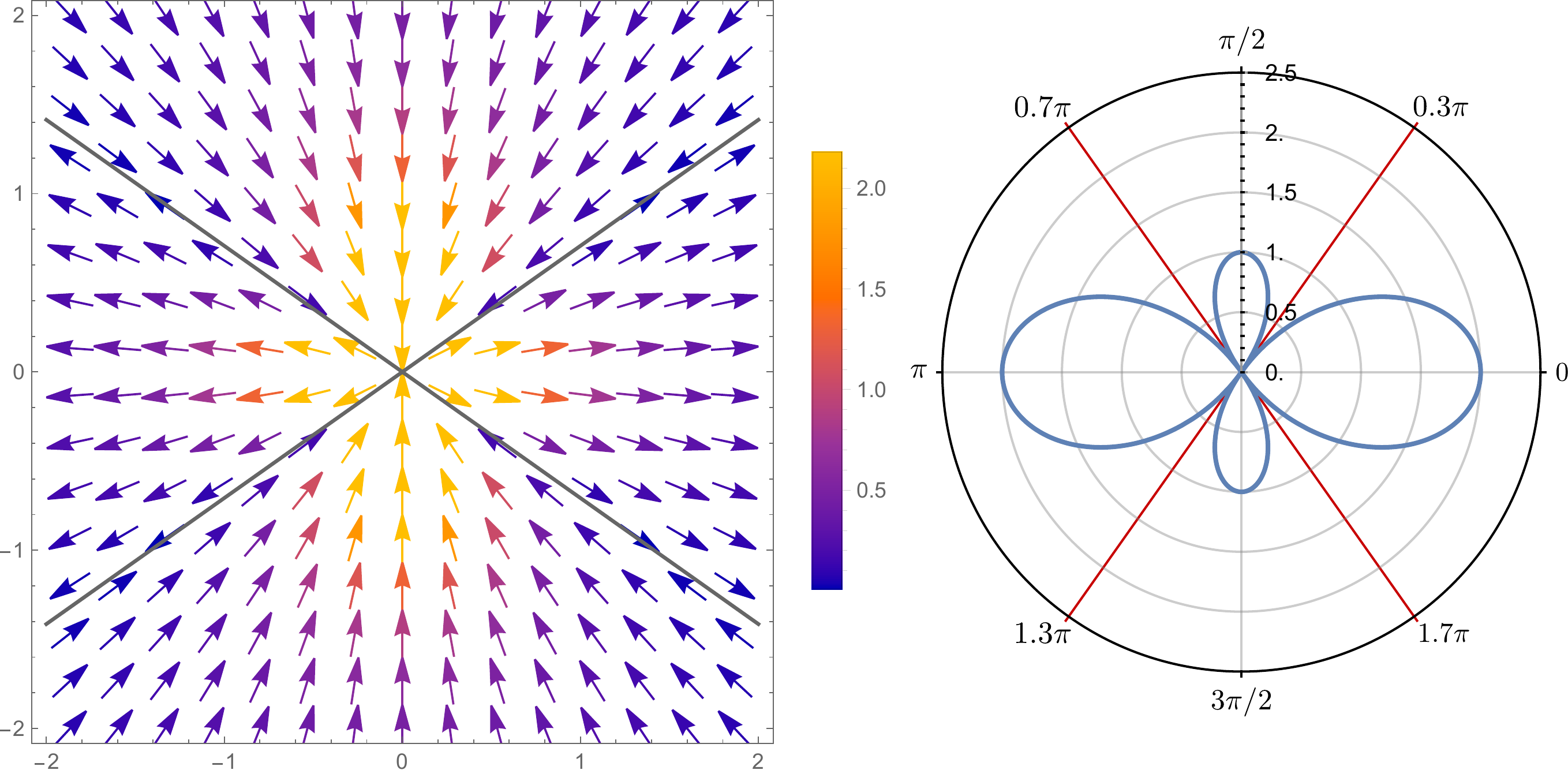}
\end{center}
\caption{Magnetic field (\protect\ref{ansa20}) produced in a constant
background by the pointlike static charge. The background fields are
directed along the vertical axis on the left panel while along the
horizontal axis on the right panel. Left: Magnetic lines of force. Owing to
the $r^{-2}$ dependence, arrow heads near the origin are bigger than those
away from the origin. Red (blue) colors indicate higher (lower) intensities
and the inclined lines mark regions of zero magnetic field. Right: Polar
plot of (\protect\ref{ansa20}). Magnetic flux is inflowing within the polar
cones $0\leq \protect\theta \leq \arccos \left( 1/\protect\sqrt{3}\right) $
and $\protect\pi -\arccos \left( 1/\protect\sqrt{3}\right) \leq \protect%
\theta \leq \protect\pi $, and outflowing within $\arccos \left( 1/\protect%
\sqrt{3}\right) \leq \protect\theta \leq \protect\pi -\arccos \left( 1/%
\protect\sqrt{3}\right) $. In the plot, $\arccos \left( 1/\protect\sqrt{3}%
\right) \approx 0.3\protect\pi$. The overall in- and
out-flows compensate each other when integrated in (\protect\ref{arb230})
with the measure $\sin {\protect\theta }d\protect\theta $ within the limits $%
(0,\protect\pi )$, thereby providing vanishing of magnetic charge.}
\label{Fig2}
\end{figure}
The field (\ref{ansa20}) is singular in one point, similar to that of a
magnetic monopole. It is radial in the sense it is parallel to the
radius-vector $\mathbf{r}$,\ but not spherically symmetric, since it depends
on the angle $\theta $\ between the radius-vector and the (common) direction
of the background fields.

The vector potential%
\begin{equation}
\mathbf{A}^{\left( 1\right) }\left( \mathbf{r}\right) =-\frac{q\mathfrak{%
\tilde{g}}}{8\pi }\frac{\left[ \mathbf{\hat{z}}\times \mathbf{\hat{r}}\right]
}{r}\cos \theta \,,  \label{stringlessA}
\end{equation}%
gives rise to the magnetic field (\ref{ansa20}) via the relation $\mathbf{B}%
^{\left( 1\right) }\left( \mathbf{r}\right) =\boldsymbol{\nabla }\times 
\mathbf{A}^{\left( 1\right) }\left( \mathbf{r}\right) $, the spatial
component of $f^{(1)\mu \nu }(\mathbf{r})$. It obeys the Coulomb gauge
condition $\mathbf{\nabla }\cdot \mathbf{A}^{\left( 1\right) }\left( \mathbf{%
r}\right) =0$. In the gauge chosen, the potential has no singularity in the
angle variable $\cos \theta =\zeta $. Therefore, nothing hinders the Bianchi
identity (\ref{bianchi}) from being fulfilled.

In order to accurately address
Eqs. (\ref{eq1}), it is possible to regularize the point charge, just as we
have done in our previous works \cite{AGS2020,AdoGitSha2016}. The solutions
presented above are consistent with the limiting cases discussed in these
references. The singular solutions discussed in the following section should
be seen as a limit of solutions for nonlinear equations with an extended
source of Coulomb field.

\section{Solutions singular on a line\label{Sec3}}

\subsection{Electric fields\label{Sec3.1}}

The Gauss (\ref{eq6.1}) and Faraday (\ref{eq11.2}) laws admit many more
solutions apart from those discussed in the preceding section. The existence
of extra solutions stems from an indeterminacy of boundary conditions to the
Cauchy problem for the set of partial differential equations (\ref{eq6.1}), (%
\ref{eq11.2}). To find these solutions, we change to spherical coordinates,
where $\mathbf{\hat{r}}$, $\boldsymbol{\hat{\theta}}$, and $\boldsymbol{\hat{%
\varphi}}$ denote, the radial, polar, and azimuthal spherical
unit vectors\footnote{%
For the angular variables, we use the convention in which the differential
volume element is $d\mathbf{r}=r^{2}\sin \theta drd\theta d\varphi $.},
respectively, and seek for solutions in the form%
\begin{equation}
\mathbf{E}\left( \mathbf{r}\right) =E_{r}\left( r,\theta \right) \mathbf{%
\hat{r}}+E_{\theta }\left( r,\theta \right) \boldsymbol{\hat{\theta}}\,.
\label{eq14.1}
\end{equation}%
Solutions of this sort exist because the induced charge density (\ref{eq10})
depends on $r$ and $\theta $, only. From now on we omit for simplicity the
superscript \textquotedblleft $\left( 1\right) $\textquotedblright\ by $%
\mathbf{E}\left( \mathbf{r}\right) $ indicating the first-order
approximation to Eqs. (\ref{eq1}) or (\ref{eq4}), since no other
approximation will be used. We only reserve this superscript as referred
specially to field (\ref{eq9.1}). In spherical coordinates, the components
obey the set of first-order partial differential equations%
\begin{eqnarray}
&&\frac{1}{r}\partial _{r}\left( r^{2}E_{r}\right) +\frac{1}{\sin \theta }%
\partial _{\theta }\left( \sin \theta E_{\theta }\right) =\frac{q\mathfrak{b}%
}{4\pi }\frac{1-3\cos ^{2}\theta }{r^{2}}\,,  \notag \\
&&\partial _{r}\left( rE_{\theta }\right) -\partial _{\theta }E_{r}=0\,,
\label{eq16}
\end{eqnarray}%
where $\partial _{r}=\partial /\partial r$, $\partial _{\theta }=\partial
/\partial \theta $. Sticking to solutions proportional to $\mathfrak{L}_{%
\mathfrak{F}}$, $\mathfrak{L}_{\mathfrak{G}}$, $\mathfrak{b}$, and $%
\mathfrak{\tilde{g}}$, electric responses (\ref{eq14.1}) must depend on $%
r^{-2}$. Hence%
\begin{equation}
E_{r}\left( r,\theta \right) =\frac{P\left( \theta \right) }{r^{2}}\,,\ \
E_{\theta }\left( r,\theta \right) =\frac{Q\left( \theta \right) }{r^{2}}\,,
\label{eq17}
\end{equation}%
whose angular parts obey the differential equations%
\begin{eqnarray}
Q^{\prime }+Q\cot \theta &=&\frac{q\mathfrak{b}}{4\pi }\left( 1-3\cos
^{2}\theta \right) \,,  \label{eq17b} \\
P^{\prime }\left( \theta \right) &=&-Q\left( \theta \right) \,.  \label{17c}
\end{eqnarray}%
Solution to the first one is%
\begin{equation}
Q\left( \theta \right) =-\frac{q\mathfrak{b}}{4\pi r^{2}}\cos \theta \sin
\theta +\frac{\tilde{\varkappa}}{r^{2}\sin \theta },  \label{Q}
\end{equation}%
where $\tilde{\varkappa}$\ is an integration constant. The part,
proportional to it in (\ref{Q}), makes an arbitrary-magnitude solution to
the equation obtained from (\ref{eq17b}) by omitting the inhomogeneity, its
right-hand side. When solving the other equation (\ref{17c}), we acquire
another integration constant $\varkappa $\ to be fixed below by a sort of
boundary condition.

Finally,\ the electric response (\ref{eq14.1}) has the general form%
\begin{equation}
\mathbf{E}\left( \mathbf{r}|\varkappa ,\tilde{\varkappa}\right) =\frac{q}{%
4\pi r^{2}}\left[ \left( \varkappa -\frac{\mathfrak{b}}{2}\cos ^{2}\theta
\right) \mathbf{\hat{r}}-\mathfrak{b}\cos \theta \sin \theta \boldsymbol{%
\hat{\theta}}\right] +\tilde{\varkappa}\boldsymbol{\Upsilon }\left( \mathbf{r%
}\right) \,,  \label{eq18}
\end{equation}%
where%
\begin{eqnarray}
&&\boldsymbol{\Upsilon }\left( \mathbf{r}\right) =\Upsilon _{r}\left(
r,\theta \right) \mathbf{\hat{r}}+\Upsilon _{\theta }\left( r,\theta \right) 
\boldsymbol{\hat{\theta}}\,,  \notag \\
&&\Upsilon _{r}\left( r,\theta \right) =-\frac{\ln \tan \theta /2}{r^{2}}%
\,,\ \ \Upsilon _{\theta }\left( r,\theta \right) =\frac{\csc \theta }{r^{2}}%
\,,\ \ 0<\theta <\pi \,.  \label{eq19a}
\end{eqnarray}%
Note that both constants $\varkappa $, $\tilde{\varkappa}$ are explicitly
shown in the argument of the electric response to distinguish (\ref{eq18})
from the field (\ref{eq9.1}). Recall that electric fields are polar vectors
under space reflections $\mathbf{r}\rightarrow -\mathbf{r}$ (or $%
r\rightarrow r$, $\theta \rightarrow \pi -\theta $, $\varphi \rightarrow
\varphi +\pi $, in spherical coordinates). This property implies that $%
\tilde{\varkappa}$ is a pseudoscalar while $\varkappa $ is a scalar (note
that\ $\cos \theta $,\ too, changes sign under reflection: $\cos \theta
\rightarrow -\cos \theta $, so does $\ln \tan \theta /2$). In the present
problem, the pseudoscalar constant is supplied by the nonlinearity to be $%
\mathfrak{\tilde{g}}$ (\ref{gtilded}) or/and $\mathfrak{L}_{\mathfrak{G}}$.
Constant $\tilde{\varkappa}$ may be thought of as a linear function of\ $%
\mathfrak{L}_{\mathfrak{G}}$\ and $\mathfrak{\tilde{g}}$ with arbirary
numerical coefficients.

The integration constant $\varkappa $ is determined to be $\varkappa =%
\mathfrak{L}_{\mathfrak{F}}+\mathfrak{b}/2$ by integrating Eq. (\ref{eq6.1})
over a spherical surface $S$\ of arbitrary radius with its center in $r=0$:%
\begin{equation}
\oint_{S}\left[ \mathbf{E}\left( \mathbf{r}|\varkappa ,\tilde{\varkappa}%
\right) -\boldsymbol{\mathfrak{E}}^{(0)}\left( \mathbf{r}\right) \right]
\cdot \mathbf{\hat{n}}dS=0\therefore \varkappa =\mathfrak{L}_{\mathfrak{F}}+%
\frac{\mathfrak{b}}{2}\,.  \label{gausse}
\end{equation}%
Therefore, we have rederived (by different method and in different
coordinates) the electric response\ (\ref{eq9.1}) and found its
generalization (\ref{eq18}), that differs from (\ref{eq9.1}) by the term
parameterized by $\tilde{\varkappa}$%
\begin{equation}
\mathbf{E}\left( \mathbf{r}|\tilde{\varkappa}\right) =\mathbf{E}^{\left(
1\right) }\left( \mathbf{r}\right) +\tilde{\varkappa}\boldsymbol{\Upsilon }%
\left( \mathbf{r}\right) \,,  \label{eq19}
\end{equation}%
that remains an unfixed integration constant, since it does not contribute
to the surface integral (\ref{gausse}) and thereof to the total charge. %
The integration constant $\tilde{\varkappa}$ } is to be chosen from among any of pseudoscalars found within the model, multiplied by arbitrary number. As long as angle-singular solutions should disappear when the background field approaches zero, in parity-even theories $\tilde{\varkappa}$ may be proportional to $\mathfrak{L}_{\mathfrak{G}}$ and/or to the combinations $\overline{\mathfrak{G}}\mathfrak{L}_{\mathfrak{FF}}$, $\overline{\mathfrak{G}}\mathfrak{L}_{\mathfrak{GG}}$, and  $\overline{\mathfrak{F}}\mathfrak{L}_{\mathfrak{FG}}$ that are dimensionally appropriate. For special parity-violating theories, this constant may depend on $\mathfrak{L}_{\mathfrak{F}}$ and/or combinations on $\overline{\mathfrak{G}}\mathfrak{L}_{\mathfrak{FG}}$ and $\overline{\mathfrak{F}}\mathfrak{L}_{\mathfrak{FF}}$, $\overline{\mathfrak{F}}\mathfrak{L}_{\mathfrak{GG}}$ , since these are now odd in pseudoscalar $\overline{\mathfrak{G}}=-\overline{\mathbf{E}}\cdot\overline{\mathbf{B}}$. Note that the first term in the right-hand-side
of (\ref{eq19}) coincides with the electric response (\ref{eq9.1}). Also,
the radial and angular components\ of $\boldsymbol{\Upsilon }\left( \mathbf{r%
}\right) $ are, in accord with (\ref{17c}), connected as%
\begin{equation}
\ln \tan \theta /2=\int_{\pi /2}^{\theta }\csc \theta ^{\prime }d\theta
^{\prime }\,.  \label{bond}
\end{equation}

The $\tilde{\varkappa}$-part\ of the electric response (\ref{eq19}), $\tilde{%
\varkappa}\boldsymbol{\Upsilon }\left( \mathbf{r}\right) $, is singular on
the $z$-axis (i.e., at $\theta =0$, $\theta =\pi $), and in the origin, $%
\mathbf{r}=\mathbf{0}$.\ Outside the singularities, the charge density
supporting this field disappears%
\begin{equation}
\boldsymbol{\nabla }\cdot \boldsymbol{\Upsilon }\left( \mathbf{r}\right)
=0\,,\ \text{ }0<\theta <\pi \,.  \label{eq19b}
\end{equation}%
Thus,\emph{\ }the field $\boldsymbol{\Upsilon }\left( \mathbf{r}\right) $ is
a free solution there and, correspondingly, it has arbirary magnitude.
Notice that the identity (\ref{eq19b}) is also valid at $\mathbf{r}=0$\
because no charge supporting the field $\boldsymbol{\Upsilon }\left( \mathbf{%
r}\right) $\ is concentrated at the origin. This fact follows from the
triviality of the Gauss theorem%
\begin{equation}
\int_{V}\boldsymbol{\nabla }\cdot \boldsymbol{\Upsilon }\left( \mathbf{r}%
\right) d\mathbf{r}=\oint_{S}\boldsymbol{\Upsilon }\left( \mathbf{r}\right)
\cdot \mathbf{\hat{n}}dS=0\,,  \label{eq19c}
\end{equation}%
applied to any sphere $S$\ of arbitrary radius centered at the singularity
point $\mathbf{r}=\mathbf{0}$\emph{.} In other words, the charge density
supporting the singular field (\ref{eq19a}) comes entirely from the $z$%
-axis, as shall be seen below.

\begin{figure}[th]
\begin{center}
\includegraphics[scale=0.33]{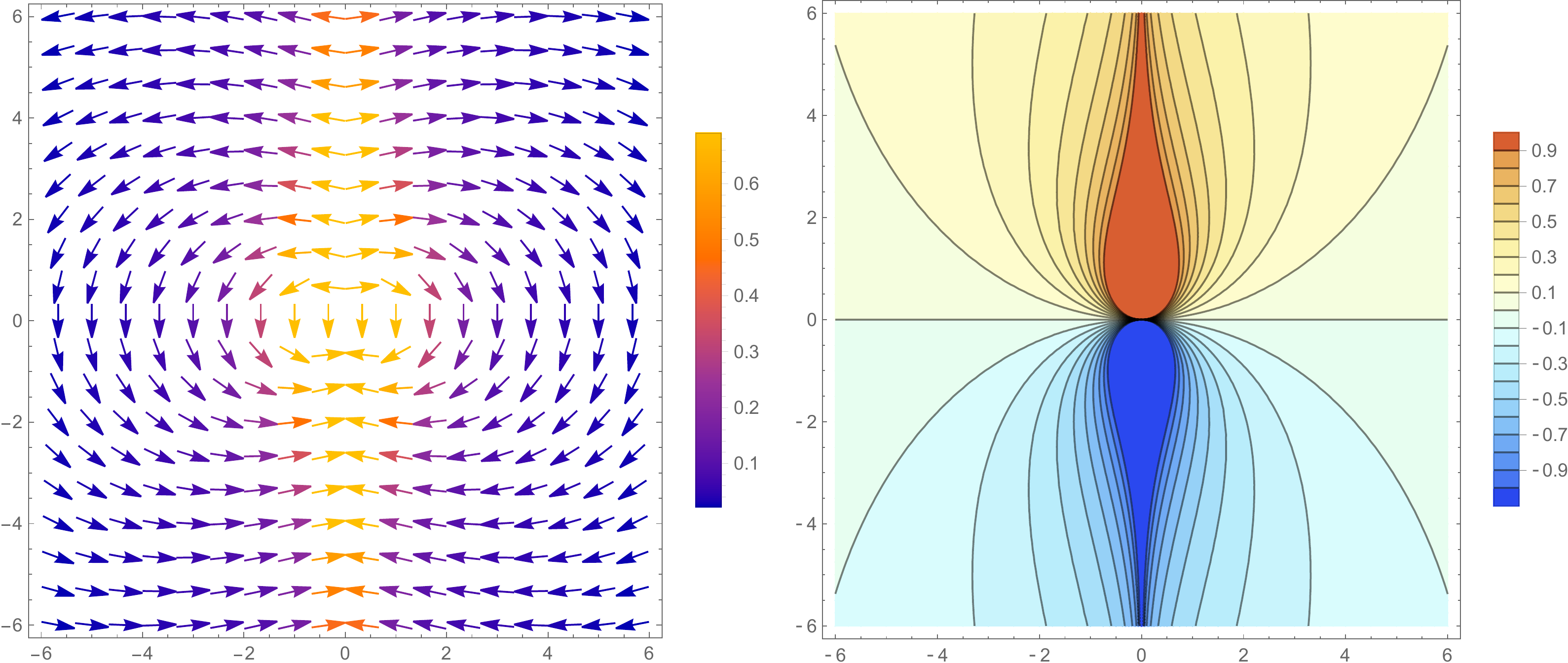}
\end{center}
\caption{Projections of the lines of force of the singular field $%
\boldsymbol{\Upsilon }\left( \mathbf{r}\right) $ (\protect\ref{eq19a}) (left
panel) and its equipotential curves $\Xi \left( \mathbf{r}\right) $ (\protect
\ref{eq21}) (solid lines, right panel) on the $yz$-plane. On the left panel,
the horizontal axis denotes the $y$-component of the field, $\Upsilon
_{y}\left( 0,y,z\right) $, while the vertical denotes the $z$-component, $%
\Upsilon _{z}\left( 0,y,z\right) $. The intensity of the field and the
potential are reflected by the color scheme dictated the labels on the right
of each plot.}
\label{Fig3}
\end{figure}

To extend the divergence (\ref{eq19b}) to the $z$-axis ($\theta =0$, $\theta
=\pi $), consider the outward flux of $\boldsymbol{\Upsilon }\left( \mathbf{r%
}\right) $ through the surface of an infinitesimally short conic cylinder
coaxial with this axis, whose arc-like bases lie at $r$ and $r+\Delta r$, $%
\Delta r\ll r$, while the side wall has its angular coordinate $\theta $.
Once $\Upsilon _{r}\left( r,\theta \right) dS_{r}=\Upsilon _{r}\left(
r,\theta \right) r^{2}\sin \theta d\theta d\varphi $ is independent of $r$
according to (\ref{eq19a}), the fluxes through the two bases cancel one
another and we are left with the flux through the wall alone%
\begin{eqnarray}
\oint_{S_{cc}}\boldsymbol{\Upsilon }\left( \mathbf{r}\right) \cdot \mathbf{%
\hat{n}}dS &=&2\pi \int_{r}^{r+\Delta r}\Upsilon _{\theta }\left( r^{\prime
},\theta \right) r^{\prime }\sin \theta dr^{\prime }  \notag \\
&=&2\pi \Delta r\Upsilon _{r}\left( r,\theta \right) r\sin \theta =2\pi 
\frac{\Delta r}{r}\,,  \label{thhh}
\end{eqnarray}%
in which Eq. (\ref{eq19a}) has been used in the last row. This relates to
the upper hemisphere ($\mathbf{\hat{r}}\geq 0$, $\boldsymbol{\hat{\theta}}%
\geq 0$). The flux (\ref{thhh}) does not depend on $\theta $, which is
expected, since the integration surface can be deformed without affecting
the flux -- unless it crosses axis $z$ -- due to the vanishing of divergence
(\ref{eq19b}). Hence, the value $\theta =0$ may be attributed to (\ref{thhh}%
). Then in the half-space of positive $z$, it holds that $r=z$, $\Delta
r=\Delta z$, and charge is distributed along positive half-axis $z$ with the
linear density 
\begin{equation}
\frac{\tilde{\varkappa}\oint_{S_{cc}}\boldsymbol{\Upsilon }\left( \mathbf{r}%
\right) \cdot \mathbf{\hat{n}}dS}{\Delta z}=\frac{2\pi \tilde{\varkappa}}{z}%
\,,  \label{linden}
\end{equation}%
inversely proportional to the distance from the seeded charge $q$. The
volume density of charge $\rho _{\mathrm{di-thr}}\left( \mathbf{r}|\tilde{%
\varkappa}\right) $ may be written as%
\begin{equation}
\tilde{\varkappa}\boldsymbol{\nabla }\cdot \boldsymbol{\Upsilon }\left( 
\mathbf{r}\right) =\frac{2\pi \tilde{\varkappa}}{z}\delta \left( \mathbf{r}%
_{\perp }\right) =\rho _{\mathrm{di-thr}}\left( \mathbf{r}|\tilde{\varkappa}%
\right) \,,\ \ \delta \left( \mathbf{r}_{\perp }\right) =\delta \left(
x\right) \delta \left( y\right) \,,  \label{exte}
\end{equation}%
so that the volume integral over the above conic cylinder, via the Gauss
theorem, be \newline
$\tilde{\varkappa}\int_{z}^{z+\Delta z}\boldsymbol{\nabla }\cdot \boldsymbol{%
\Upsilon }\left( \mathbf{r}\right) dz\int dxdy=2\pi \tilde{\varkappa}\Delta
z/z$ in agreement with the flux (\ref{linden}). The factor $%
2\pi $ may be omitted in view of arbitrariness of $\tilde{\varkappa}$. It
can be shown that Eq. (\ref{exte}) retains its form at negative $z$ as well,
which fact is in accord with the nullification of the total charge in the
whole space, explained above (after eq. (\ref{eq19a})). Therefore, the signs
of the charge of the thread are opposite at opposite sides of the point $z=0$%
. This suggests the name \textquotedblleft di-thread\textquotedblright\ by
analogy with \textquotedblleft dipole\textquotedblright . The pattern of
lines of force is shown in Fig. \ref{Fig3}. These are formed by arcs,
starting from the positive $z$-axis and ending on the negative $z$-axis.
Note that, once $1/z$ reverses sign under space reflection, the charge
density (\ref{exte}) is a scalar function in agreement with the condition
that the divergence of electric field must result in a scalar function.

To conclude, Eq. (\ref{exte}) is our promised extension of relation (\ref%
{eq19b}) to include axis $z$. The background of parallel external constant
magnetic and electric fields perturbed by a seeded point charge predisposes
to considering exotic charged di-thread and electric field produced by it by
supplying a pseudoscalar constant necessary to form the charge -- the same
as to form magnetic charge in the next Section and, generally, in nature as
well.

The issue is quite similar to the standard point-like charge $Q$ in
Maxwell's electrodynamics.\ Outside the origin $\mathbf{r}=\mathbf{0}$,
where the point charge is located,\ its electric field obeys the free
equation $\boldsymbol{\nabla }\cdot \mathbf{E}\left( \mathbf{r}\right) =0$,\
whose solution (decreasing in the far-off domain) is $\mathbf{E}\left( 
\mathbf{r}\right) =\left( Q/4\pi r^{3}\right) \mathbf{r}$,\ where $Q$\
arises as an arbitrary integration constant to be identified with the charge
that supports the field. Its role is analogous to that of our integration
constant $\tilde{\varkappa}$\ that determines the magnitude of the charge of
the di-thread (\ref{exte}).

Similar to the point charge $Q$, the charge density of the di-thread (\ref%
{exte}) should be understood as an externally given parameter, in contrast
to the induced charge density (\ref{eq10}) which depends exclusively on the
seeded charge $q$. Correspondingly, the thread density \ $A^{0}\left( 
\mathbf{r}\right) \rho _{\mathrm{di-thr}}\left( \mathbf{r}|\tilde{\varkappa}%
\right) $\ should be added to Lagrangian in the action (\ref{L}) and appear
in the right-hand side of the equation in place of (\ref{eq6.1}); cf.
analogous statement concerning Dirac string current in the next Section.

To conclude this section, we emphasize that the Bianchi identity%
\begin{equation}
\boldsymbol{\nabla }\times \boldsymbol{\Upsilon }\left( \mathbf{r}\right) =%
\mathbf{0}\,,  \label{eq21a}
\end{equation}%
holds everywhere, including at the singularities $\mathbf{r}=\mathbf{0}$, $%
\theta =0$, and $\theta =\pi $. This result follows from the triviality of
the Stokes theorem%
\begin{equation}
\int_{\mathcal{S}}\left[ \boldsymbol{\nabla }\times \boldsymbol{\Upsilon }%
\left( \mathbf{r}\right) \right] \cdot \mathbf{\hat{n}}d\mathcal{S}=\oint_{%
\mathcal{C}}\boldsymbol{\Upsilon }\left( \mathbf{r}\right) \cdot d\mathbf{l}%
=0\,,  \label{eq21c}
\end{equation}%
applied to any closed contour \emph{$\mathcal{C}$}$\ $enclosing the origin
and any segment of the $z$-axis. To reach this conclusion, we calculated the
integral on the right for two different contours. The first was a circle of
arbitrary radius, centered at the $z$-axis and perpendicular to it. This
circle could be placed at any height with respect to the $xy$-plane. The
second contour was a rectangle of any shape, located at a plane with $%
\mathbf{r}_{\perp }=\mathrm{const.}$, enclosing the origin (but not passing
through it) and a segment of the $z$-axis. In the first case, the line
integral in (\ref{eq21c}) is identically zero because the field $\boldsymbol{%
\Upsilon }\left( \mathbf{r}\right) $ is always perpendicular to the unit
vector tangent to the circle $d\mathbf{l}=\left\vert \mathbf{r}_{\perp
}\right\vert d\varphi \boldsymbol{\hat{\varphi}}$\ , i.e., $\boldsymbol{%
\Upsilon }\left( \mathbf{r}\right) \cdot d\mathbf{l}=0$, where $\left\vert 
\mathbf{r}_{\perp }\right\vert =\sqrt{x^{2}+y^{2}}$ denotes the magnitude of
the polar radius and $\varphi \in \left[ 0,2\pi \right) $ the polar angle.
As for the rectangular contour, the line integral vanishes identically,
regardless the size of the rectangle. This identity can be straightforwardly
shown using the representation of\emph{\ }$\boldsymbol{\Upsilon }\left( 
\mathbf{r}\right) $\ in Cartesian coordinates,%
\begin{eqnarray}
\boldsymbol{\Upsilon }\left( \mathbf{r}\right) &=&\Upsilon _{\perp }\left( 
\mathbf{r}\right) \mathbf{\hat{r}}_{\perp }+\Upsilon _{\parallel }\left( 
\mathbf{r}\right) \mathbf{\hat{z}}\,,\ \ \mathbf{\hat{r}}_{\perp }=\frac{x%
\mathbf{\hat{x}}+y\mathbf{\hat{y}}}{\left\vert \mathbf{r}_{\perp
}\right\vert }\,,  \notag \\
\Upsilon _{\perp }\left( \mathbf{r}\right) &=&\frac{1}{\mathbf{r}_{\perp
}^{2}+z^{2}}\left( \frac{z}{\left\vert \mathbf{r}_{\perp }\right\vert }-%
\frac{\left\vert \mathbf{r}_{\perp }\right\vert }{2\sqrt{\mathbf{r}_{\perp
}^{2}+z^{2}}}\ln \frac{\sqrt{\mathbf{r}_{\perp }^{2}+z^{2}}-z}{\sqrt{\mathbf{%
r}_{\perp }^{2}+z^{2}}+z}\right) \,,  \notag \\
\Upsilon _{\parallel }\left( \mathbf{r}\right) &=&-\frac{1}{\mathbf{r}%
_{\perp }^{2}+z^{2}}\left( 1+\frac{z}{2\sqrt{\mathbf{r}_{\perp }^{2}+z^{2}}}%
\ln \frac{\sqrt{\mathbf{r}_{\perp }^{2}+z^{2}}-z}{\sqrt{\mathbf{r}_{\perp
}^{2}+z^{2}}+z}\right) \,.  \label{eq21d}
\end{eqnarray}%
As a result, no current flux through the origin and the z-axis exists, which
means that the singular field is a pure gradient field%
\begin{equation}
\boldsymbol{\Upsilon }\left( \mathbf{r}\right) =-\boldsymbol{\nabla }\Xi
\left( \mathbf{r}\right) \,,\ \ \Xi \left( \mathbf{r}\right) =-\frac{\ln
\tan \theta /2}{r}+\tilde{\xi}\,,  \label{eq21}
\end{equation}%
everywhere. Here, $\tilde{\xi}$ is a pseudoscalar constant. In other words,
the electric field (\ref{eq19}) respects the Bianchi identity (\ref{eq8}) at
all points,\emph{\ }$\boldsymbol{\nabla }\times \mathbf{E}\left( \mathbf{r}|%
\tilde{\varkappa}\right) =\mathbf{0}$.\ Notice that the potential $\Xi
\left( \mathbf{r}\right) $\ is a pseudoscalar function, as it should be
since $\boldsymbol{\Upsilon }\left( \mathbf{r}\right) $\ is a
space-reflection invariant vector field.\emph{\ }Thus, up to an unimportant
constant, the singular part of $\mathbf{E}^{\left( 1\right) }\left( \mathbf{r%
}|\tilde{\varkappa}\right) $ can be written as the gradient of the scalar
potential%
\begin{equation}
A_{0}\left( \mathbf{r}|\tilde{\varkappa}\right) =A_{0}^{\left( 1\right)
}\left( \mathbf{r}\right) +\tilde{\varkappa}\Xi \left( \mathbf{r}\right) \,,
\label{eq21b}
\end{equation}%
where the first term is given by Eq. (\ref{eq11b}). It should be noted that
just like the singular field $\tilde{\varkappa}\boldsymbol{\Upsilon }\left( 
\mathbf{r}\right) $, the scalar potential $\tilde{\varkappa}\Xi \left( 
\mathbf{r}\right) $is also singular at the $z$-axis, the origin $\mathbf{r}=%
\mathbf{0}$ included. 

\subsection{Magnetic fields\label{Ss3.2}}

In contrast to the magnetically neutral magnetic field (\ref{ansa20})
created in a constant two-field background by a small point-like Coulomb
source, we discuss below a magnetic solution to the same field equation,
which carries a nontrivial magnetic charge. Similarly to the electric
solutions, the linearized Maxwell's equations for magnetic ones (\ref{nnew7b}%
) admit more solutions apart from those considered in Sec. \ref{Sec2}.
Firstly, we seek for solutions in the non-radial, angle-dependent form%
\begin{equation}
\mathbf{B}\left( \mathbf{r}\right) =B_{r}\left( r,\theta \right) \mathbf{%
\hat{r}}+B_{\theta }\left( r,\theta \right) \boldsymbol{\hat{\theta}}\,,
\label{eq14.2}
\end{equation}%
and rewrite the Maxwell equations (\ref{nnew7b}), (\ref{bianchi}) in
spherical coordinates to learn that the components obey the set of equations%
\begin{eqnarray}
&&\partial _{r}\left( rB_{\theta }\right) -\partial _{\theta }B_{r}=-\frac{3%
\mathfrak{\tilde{g}}q}{4\pi }\frac{\cos \theta \sin \theta }{r^{2}}\,, 
\notag \\
&&\partial _{\theta }\left( \sin \theta B_{\theta }\right) +\frac{1}{r}%
\partial _{r}\left( r^{2}B_{r}\right) =0\,.  \label{eq22}
\end{eqnarray}%
Notice that we omitted the superscript \textquotedblleft $\left( 1\right) $%
\textquotedblright\ by $\mathbf{B}\left( \mathbf{r}\right) $ in (\ref{eq14.2}%
) to avoid confusion with the magnetic response (\ref{ansa20}). We adopt
this convention from now on.

Restricting to magnetic solutions proportional to $\mathfrak{L}_{\mathfrak{F}%
}$, $\mathfrak{L}_{\mathfrak{G}}$, $\mathfrak{b}$, and $\mathfrak{\tilde{g}}$%
, the fields (\ref{eq14.2}) must depend on the radial coordinate as $r^{-2}$,%
\begin{equation}
B_{r}\left( r,\theta \right) =\frac{J\left( \theta \right) }{r^{2}}\,,\ \
B_{\theta }\left( r,\theta \right) =\frac{K\left( \theta \right) }{r^{2}}\,.
\label{eq23}
\end{equation}%
Then, plugging the ansatzes (\ref{eq23}) into Eqs. (\ref{eq22}) we find%
\begin{equation}
\partial _{\theta }\left( K\sin \theta \right) =0\,,\ \ J^{\prime }+K=\frac{%
3q\mathfrak{\tilde{g}}}{4\pi }\cos \theta \sin \theta \,.  \label{eq24}
\end{equation}%
Solving the first equation in (\ref{eq24}) supplies the angular component
with an arbitrary constant: $K\left( \theta \right) =\kappa \csc \theta $, $%
\kappa =\mathrm{const}$. Plugging this solution into the equation for $%
J\left( \theta \right) $ we finally obtain%
\begin{equation}
\mathbf{B}\left( \mathbf{r}|\tilde{\kappa},\kappa \right) =\mathbf{B}_{%
\mathrm{ch}}\left( \mathbf{r}|\tilde{\kappa}\right) +\kappa \boldsymbol{%
\Upsilon }\left( \mathbf{r}\right) \,,  \label{eq25}
\end{equation}%
where $\tilde{\kappa}$ is another integration constant, stemming from the
equation for $J\left( \theta \right) $ in (\ref{eq24}), the vector field $%
\boldsymbol{\Upsilon }\left( \mathbf{r}\right) $ was defined before (\ref%
{eq19a}), and%
\begin{equation}
\mathbf{B}_{\mathrm{ch}}\left( \mathbf{r}|\tilde{\kappa}\right) =\left( 
\tilde{\kappa}-\frac{3q\mathfrak{\tilde{g}}}{8\pi }\cos ^{2}\theta \right) 
\frac{\mathbf{\hat{r}}}{r^{2}}\,.  \label{bcharged}
\end{equation}%
Following the same arguments that led to Eq. (\ref{eq19}), i.e., appealing
to the fact that magnetic fields are axial vectors, we conclude that $\tilde{%
\kappa}$ is a pseudoscalar while $\kappa $ is a scalar. Magnetic field (\ref%
{eq25}) contains the response (\ref{ansa20}) in itself, bearing in mind that 
$\tilde{\kappa}$\ is arbitrary. $\mathbf{B}_{\mathrm{ch}}\left( \mathbf{r}|%
\tilde{\kappa}\right) $\ is radial, whereas $\kappa \boldsymbol{\Upsilon }%
\left( \mathbf{r}\right) $ depends on $\theta $. Notice that we regain the
magnetic response (\ref{ansa20}) from (\ref{bcharged}) by choosing $\tilde{%
\kappa}=q\mathfrak{\tilde{g}}/8\pi $; i.e., $\mathbf{B}_{\mathrm{ch}}\left( 
\mathbf{r}|q\mathfrak{\tilde{g}}/8\pi \right) =\mathbf{B}^{\left( 1\right)
}\left( \mathbf{r}\right) $.

Unlike the electric solutions (\ref{eq19}), magnetic solutions found above
are twice arbitrary because both constants remain undetermined by the
consistency with the Bianchi identity (\ref{bianchi}) and the induced
current density Eq. (\ref{par}). It should be noted that $\boldsymbol{\nabla 
}\times \mathbf{B}\left( \mathbf{r}|\tilde{\kappa},\kappa \right) =%
\boldsymbol{\nabla }\times \mathbf{B}^{(1)}\left( \mathbf{r}\right) $\
beyond the singularities, hence the solutions (\ref{eq25}) may be
alternatively presented as%
\begin{equation}
\mathbf{B}\left( \mathbf{r}|\tilde{\kappa},\kappa \right) =\mathbf{B}%
^{\left( 1\right) }\left( \mathbf{r}\right) +\boldsymbol{\nabla }\Omega
\left( \mathbf{r}|\tilde{\kappa}\right) -\kappa \boldsymbol{\nabla }\Xi
\left( \mathbf{r}\right) \,,  \label{arb29}
\end{equation}%
where $\Omega \left( \mathbf{r}|\tilde{\kappa}\right) $ is the pseudoscalar
potential%
\begin{equation*}
\Omega \left( \mathbf{r}|\tilde{\kappa}\right) =-\frac{1}{r}\left( \tilde{%
\kappa}-\frac{q\mathfrak{\tilde{g}}}{8\pi }\right) +\tilde{\omega}\,,
\end{equation*}%
$\tilde{\omega}$ is a pseudoscalar constant, and $\Xi \left( \mathbf{r}%
\right) $ is defined in Eq. (\ref{eq21}).

We shall be interested first in the radial solutions of equation (\ref{eq25}%
), for which the pseudoscalar constant $\tilde{\kappa}$\ may be formed by
parameters of our problem: (dimensionally consistent) pseudoscalar
combinations of the field invariants $\overline{\mathfrak{G}}$, $\overline{%
\mathfrak{F}}$, the derivatives $\mathfrak{L}_{\mathfrak{G}}$, $\mathfrak{L}%
_{\mathfrak{FG}}$, included. It must be noted that usual electrodynamics in
the vacuum does not provide such a constant pseudoscalar. The Bianchi
identity (\ref{bianchi}) 
\begin{subequations}
\begin{equation}
\boldsymbol{\nabla }\cdot \mathbf{B}_{\mathrm{ch}}\left( \mathbf{r}|\tilde{%
\kappa}\right) =0\,,\ \ \mathbf{r}\neq \mathbf{0}\,,  \label{arb31}
\end{equation}%
is fulfilled everywhere, but not at the origin $\mathbf{r}=\mathbf{0}$. In
other words, the magnetic charge is concentrated in a point. Its value is
defined by the surface integral 
\end{subequations}
\begin{equation}
q_{M}=\oint_{S}\left( \mathbf{B}_{\mathrm{ch}}\left( \mathbf{r}|\tilde{\kappa%
}\right) \cdot \mathbf{\hat{n}}\right) dS=4\pi \tilde{\kappa}-\frac{q%
\mathfrak{\tilde{g}}}{2}\,,  \label{s3.2}
\end{equation}%
wherein the surface $S$ embracing the charge may be chosen\ as a sphere of
arbitrary radius.

Following Gauss' theorem $\oint_{S}\left( \mathbf{B}_{\mathrm{ch}}\left( 
\mathbf{r}|\tilde{\kappa}\right) \cdot \mathbf{\hat{n}}\right) dS=\int_{V}%
\left[ \boldsymbol{\nabla }\cdot \mathbf{B}_{\mathrm{ch}}\left( \mathbf{r}|%
\tilde{\kappa}\right) \right] d^{3}x$, one unites (\ref{arb31}) and (\ref%
{s3.2}) in the form%
\begin{equation}
\boldsymbol{\nabla }\cdot \mathbf{B}_{\mathrm{ch}}\left( \mathbf{r}|\tilde{%
\kappa}\right) =q_{M}\delta \left( \mathbf{r}\right) \,.  \label{magcharge}
\end{equation}%
It is noteworthy that $\tilde{\kappa}$ is not arbitrary, as the magnetic charge must vanish in the limit of zero background. This requirement imposes certain restrictions on $\tilde{\kappa}$. For example, this constant could depend on $\mathfrak{L}_{\mathfrak{G}}$ and/or dimensionally consistent combinations of $\overline{\mathfrak{F}}$ with $\mathfrak{L}_{\mathfrak{FG}}$ and of $\overline{\mathfrak{G}}$ with $\mathfrak{L}_{\mathfrak{FF}}$, $\mathfrak{L}_{\mathfrak{GG}}$.
In the special case of a P-odd theory,\ the magnetic
charge may exist with no electric field in the background. Indeed, with $%
\overline{\mathbf{E}}=\overline{\mathfrak{G}}=0$, one finds that the
remaining coefficients $\mathfrak{L}_{\mathfrak{FG}}$\ and $\mathfrak{L}_{%
\mathfrak{G}}$ are zero, unless the Lagrangian itself contains a
parity-violating term $\chi \left( \mathfrak{F}\right) \overline{\mathfrak{G}%
}$ linear in $\overline{\mathfrak{G}}$,\ so that $\mathfrak{L}_{\mathfrak{FG}%
}=\chi ^{\prime }\left( \mathfrak{F}\right) \neq 0$, $\mathfrak{L}_{%
\mathfrak{G}}=\chi \left( \mathfrak{F}\right) \neq 0$. Moreover, in the
limit of small constant fields, when we must set $\overline{\mathbf{E}}=%
\overline{\mathbf{B}}=0$ inside the coefficient functions, one has in a
parity-even theory $\mathfrak{L}_{\mathfrak{G}}=\mathfrak{L}_{\mathfrak{FG}%
}=0$. In this case, Eq. (\ref{ansa20}) (or Eq. (\ref{bcharged}), with $%
\tilde{\kappa}=q\mathfrak{\tilde{g}}/8\pi $) coincides with the solution
numbered as (12) in our previous paper \cite{AdoGitSha2015} taken in the
limit $b=0$. Note that the approximation exploited in \cite{AdoGitSha2015}
is different from that in the present paper, as commented above Eq. (\ref%
{eq4}).

It remains to consider the last portion of the magnetic solution (\ref{eq25}%
), proportional to $\kappa $. According to the discussion in the preceding
subsection, the singular field $\mathbf{\Upsilon }\left( \mathbf{r}\right) $
is supported by a charge density concentrated on the $z$-axis and
symmetrically distributed with positive/negative signs along the
positive/negative semi-axes (\ref{exte}). This means that in addition to the
magnetic charge (\ref{s3.2}), the magnetic field (\ref{eq25}) is supported
by a\textit{\ magnetic di-thread} -- a generalization of a magnetic dipole
-- owing to $\kappa \mathbf{\Upsilon }\left( \mathbf{r}\right) $, whose
density reads%
\begin{equation}
\boldsymbol{\nabla }\cdot \left[ \kappa \mathbf{\Upsilon }\left( \mathbf{r}%
\right) \right] =\frac{2\pi \kappa }{z}\delta \left( \mathbf{r}_{\perp
}\right) \,.  \label{magthread}
\end{equation}%
This magnetic charge density is understood as an external agent to be
included into the primary Lagrangian. Like the magnetic charge (\ref%
{magcharge}), the above charge density is a pseudoscalar function, as it
should be since the magnetic field $\kappa \mathbf{\Upsilon }\left( \mathbf{r%
}\right) $ is an axial vector field, as seen from (\ref{eq19a}). %
It is important to note that the value of $\kappa$ is not
arbitrary, since the non-radial singular part of the solution (\ref{arb29})
should disappear when the background field approaches zero. If the
nonlinearity of the theory is symmetric under space reflection, then
$\kappa$ may include dimensionless parity-even products of
$\overline{\mathfrak{G}}$ with $\mathfrak{L}_{\mathfrak{FG}}$ and of $\overline{\mathfrak{F}}$ with
$\mathfrak{L}_{\mathfrak{FF}}$, $\mathfrak{L}_{\mathfrak{GG}}$. As for
parity-violating theories, $\kappa$ may depend on dimensionless parity-even products of $\overline{\mathfrak{G}}$ with $\mathfrak{L}_{\mathfrak{FF}}$,
$\mathfrak{L}_{\mathfrak{GG}}$ and of $\overline{\mathfrak{F}}$ with
$\mathfrak{L}_{\mathfrak{FG}}$. The magnetic field lines are the same as
the electric lines depicted in the left panel of Fig. \ref{Fig3}.

\subsubsection{Singular vector potential and the Dirac string\label{Ss4.3}}

The vector potential%
\begin{equation}
\mathbf{A}_{\mathrm{ch}}^{\left( \pm \right) }\left( \mathbf{r}|\tilde{\kappa%
}\right) =\frac{\sin \theta }{4\pi r}\left( -\frac{q\mathfrak{\tilde{g}}}{2}%
\cos \theta \pm \frac{q_{M}}{1\pm \cos \theta }\right) \boldsymbol{\hat{%
\varphi}}\,,  \label{4.12}
\end{equation}%
which satisfies the equation to be obtained from (\ref{nnew7b}) by the
formal substitution \newline
$\mathbf{B}_{\mathrm{ch}}\left( \mathbf{r}|\tilde{\kappa}\right) =%
\boldsymbol{\nabla }\times \mathbf{A}_{\mathrm{ch}}^{\left( \pm \right)
}\left( \mathbf{r}|\tilde{\kappa}\right) $, is singular on the
negative/positive half-axis $\cos \theta =\mp 1$ drawn along the common
direction $\mathbf{\hat{z}}$ of the background fields. The Coulomb gauge
condition 
\begin{equation}
\mathbf{\nabla }\cdot \mathbf{A}_{\mathrm{ch}}^{\left( \pm \right) }\left( 
\mathbf{r}|\tilde{\kappa}\right) =0,\text{ \ }\zeta \neq \mp 1\,,\ \ \mathbf{%
r}\neq \mathbf{0}\,,  \label{Coulomb}
\end{equation}%
is satisfied by (\ref{4.12}) beyond the singularity line. Its extension to
the singularity line leaves it as it is. Indeed, the flux $\oint_{\mathcal{S}%
}\mathbf{A}_{\mathrm{ch}}^{\left( \pm \right) }\left( \mathbf{r}|\tilde{%
\kappa}\right) \cdot \mathbf{\hat{r}}d\mathcal{S}$\ through a sphere $%
\mathcal{S}$ of radius $r$ is identically zero. The reason is that the
normal vector pointing out the sphere is just the radius vector $\mathbf{%
\hat{r}}$,\ which is orthogonal to $\boldsymbol{\hat{\varphi}}$, namely $%
\boldsymbol{\hat{\varphi}}\cdot \mathbf{\hat{r}}=0$. Analogously, the flux
of $\mathbf{A}_{\mathrm{ch}}^{\left( \pm \right) }\left( \mathbf{r}|\tilde{%
\kappa}\right) $ through an infinitely long cylinder surrounding the lines
of singularities is zero too, since (\ref{4.12}) is tangent both to its
walls and its base.

The potential (\ref{4.12}) produces (\ref{bcharged}) outside the singular
half-axes%
\begin{equation}
\mathbf{B}_{\mathrm{ch}}\left( \mathbf{r}|\tilde{\kappa}\right) =\boldsymbol{%
\nabla }\times \mathbf{A}_{\mathrm{ch}}^{\left( \pm \right) }\left( \mathbf{r%
}|\tilde{\kappa}\right) \,,\ \text{ }\mathbf{r}\neq \mathbf{0},\ \text{ }%
\zeta \neq \mp 1\,,  \label{CurlAGerman}
\end{equation}%
but at the region of singularities it gives rise to singular magnetic fields
-- \textquotedblleft string\textquotedblright\ magnetic fields\textrm{\ }$%
\mathbf{B}_{\mathrm{string}}^{\mp }\left( \mathbf{r}|\tilde{\kappa}\right) $%
\textrm{. }To compute these contributions, we may extend the above relation
to the forbidden half-axis through Stokes's theorem (recall that we choose
the axis $z$ along the background field; the axes $x$ and $y$ make Cartesian
coordinates in the orthogonal plane $\mathbf{r}_{\perp }=(x,y)$)%
\begin{equation}
\oint_{\mathcal{C}_{\mp }}\mathbf{A}_{\mathrm{ch}}^{\left( \pm \right)
}\left( \mathbf{r}|\tilde{\kappa}\right) \cdot d\mathbf{l}=\int_{\mathcal{S}%
_{\mp }}\left( \mathbf{B}_{\mathrm{string}}^{\mp }\left( \mathbf{r}|\tilde{%
\kappa}\right) \cdot \mathbf{\hat{z}}\right) dxdy\,,  \label{stokes}
\end{equation}%
where the integral on the left is taken along any of circles $\mathcal{C}%
_{\mp }$ (enclosing the corresponding open surfaces $\mathcal{S}_{\mp }$)
with a vanishing radius, $\left\vert \mathbf{r}_{\perp }\right\vert =r\sin
\theta =r\sqrt{1-\zeta ^{2}}\rightarrow 0$ as $\zeta \rightarrow \mp 1$,
about the half-axis of singularities $z\lessgtr 0$. Any of the two circles $%
\mathcal{C}_{\mp }$ is a section of a plane\ orthogonal to $z$ with the
spherical surface, whose radius is $r$. The surface integral on the right
may be taken over the part $\mathcal{S}_{\mp }$ of the orthogonal plane
restricted by the said circle. The left-hand integral is calculated as
follows (the part, nonsingular on the half-axis, does not contribute in the
limit $\zeta \rightarrow \mp 1$):%
\begin{equation}
\oint_{\mathcal{C}_{\mp }}\mathbf{A}_{\mathrm{ch}}^{\left( \pm \right)
}\left( \mathbf{r}|\tilde{\kappa}\right) \cdot d\mathbf{l}=\frac{1}{4\pi }%
\oint_{\mathcal{C}_{\mp }}\frac{\sin \theta }{r}\left( -\frac{q\mathfrak{%
\tilde{g}}}{2}\cos \theta \pm \frac{q_{M}}{1\pm \cos \theta }\right) 
\boldsymbol{\hat{\varphi}}\cdot d\mathbf{l}=\pm q_{M}\,.  \label{circulation}
\end{equation}

Eqs.(\ref{circulation}), (\ref{stokes}) establish the value of the flux of
the extra part of the magnetic field concentrated on singular half-axes
through the orthogonal plane (cf. the procedure in \cite{Heras18}, \cite%
{Yakov} described for the Dirac monopole \cite{Dirac1}; see also the review 
\cite{review}. Significance of the string field was also revealed in \cite%
{Dunia2}). To satisfy Eqs. (\ref{circulation}), (\ref{stokes}) we have to
prescribe the following distributional expression to that \textquotedblleft
string\textquotedblright\ magnetic field. It has only one, $z$-th,
component, and its magnitude is%
\begin{equation}
\mathbf{B}_{\mathrm{string}}^{\mp }\left( \mathbf{r}|\tilde{\kappa}\right)
=\pm q_{M}\Theta \left( \mp z\right) \delta \left( \mathbf{r}_{\perp
}\right) \mathbf{\hat{z}\,},  \label{Bstring}
\end{equation}%
where $\Theta $ is the Heaviside step function. Finally, Eq.
(\ref{CurlAGerman}) is extended to the singular half-axis as%
\begin{equation*}
\mathbf{B}_{\mathrm{ch}}\left( \mathbf{r}|\tilde{\kappa}\right) =\boldsymbol{%
\nabla }\times \mathbf{A}_{\mathrm{ch}}^{\left( \pm \right) }\left( \mathbf{r%
}|\tilde{\kappa}\right) =\mathbf{B}_{\mathrm{ch}}\left( \mathbf{r}|\tilde{%
\kappa}\right) +\mathbf{B}_{\mathrm{string}}^{\mp }\left( \mathbf{r}|\tilde{%
\kappa}\right) =\mathbf{B}_{\mathrm{ch}}\left( \mathbf{r}|\tilde{\kappa}%
\right) \pm q_{M}\Theta \left( \mp z\right) \delta \left( \mathbf{r}_{\perp
}\right) \mathbf{\hat{z}}\,.
\end{equation*}

Contrary to the Dirac monopole, in our case the monopole solution (\ref%
{bcharged}) is not center-symmetric. P.A.-M. Dirac \cite{Dirac2} came to his intriguing prediction about discreteness of the
 product of electric and magnetic charges, pursuing the goal of making the string invisible for electrons. That 
was an obligation imposed by gauge invariance, because gauge transformation is able to alter configuration
 of the string. After the Aharonov-Bohm effect \cite{Aharonov} had been discovered, the string invisibility
 became questionable. In our context, the string appears correlated with the direction of the external field. The
 situation may be viewed upon as a spontaneous breakdown: the string direction may be any, but is fixed by
 an external agent, the background field. As long as the gauge symmetry has been broken, we are no longer
 free to change the string orientation by a gauge transformation, and we see no reason for invisibility of string 
and charge discretization either. Moreover, in the next Subsubsection we shall see that the string field
 (\ref{Bstring}) requires a solenoidal current (\ref{sb13}) as its source, axis of solenoid being the
 string direction. Certainly, current is not a subject of gauge transformation. This means that to give a magnetic
 monopole is to give its place, its charge and (direction of the axis of) its string current. If we choose this 
direction differently, we define a different monopole. This remark does not exclude the
possible achievement of charge quantization by considering discreteness of
angular momentum of electromagnetic fields. Such endeavors are attracting
attention, for analysis with the account of the Dirac string see \cite{Dunia}.

\subsubsection{String current\label{Ss4.4}}

The usual trouble characteristic of the standard magnetic monopole of
Maxwell electrodynamics cured only by passing to nonAbelian gauge theory 
\cite{Polyakov}, (see also \cite{Yakov}, \cite{Rossi}) are shared by our
magnetic monopole (except that ours comes in already equipped with the
necessary pseudoscalar coefficient).

Indeed, the full magnetic field produced by vector-potential (\ref{4.12})%
\begin{equation}
\mathbf{B}_{\mathrm{string}}^{\mp }\left( \mathbf{r}|\tilde{\kappa}\right) +%
\mathbf{B}_{\mathrm{ch}}\left( \mathbf{r}|\tilde{\kappa}\right) \,,
\label{not-solution}
\end{equation}%
is not a solution to Eq. (\ref{nnew7b}). This fact follows from the
nontriviality of the current flowing around the infinitely thin cylinder%
\begin{equation}
\mathbf{j}_{\mathrm{string}}^{\mp }\left( \mathbf{r}|\tilde{\kappa}\right) =%
\boldsymbol{\nabla }\times \mathbf{B}_{\mathrm{string}}^{\mp }\left( \mathbf{%
r}|\tilde{\kappa}\right) =\pm q_{M}\left[ \mathbf{\hat{x}}\delta \left(
x\right) \delta ^{\prime }\left( y\right) -\mathbf{\hat{y}}\delta ^{\prime
}\left( x\right) \delta \left( y\right) \right] \Theta \left( \mp z\right)
\,,  \label{sb13}
\end{equation}%
which is not compatible with the initial set of nonlinear equations (\ref%
{nnew7b}). More precisely, by substituting (\ref{not-solution}) in place of $%
\mathbf{B}^{\left( 1\right) }\left( \mathbf{r}\right) $ into Eq. (\ref%
{nnew7b}) (with $\mathbf{j}^{\left( 1\right) }\left( \mathbf{r}\right) $
defined by (\ref{par})) 
\begin{equation*}
\boldsymbol{\nabla }\times \mathbf{B}^{\left( 1\right) }\left( \mathbf{r}%
\right) =\boldsymbol{\nabla }\times \left[ \mathbf{B}_{\mathrm{string}}^{\mp
}\left( \mathbf{r}|\tilde{\kappa}\right) +\mathbf{B}_{\mathrm{ch}}\left( 
\mathbf{r}|\tilde{\kappa}\right) \right] =\mathbf{j}^{\left( 1\right)
}\left( \mathbf{r}\right) \,,
\end{equation*}%
and bearing in mind that $\boldsymbol{\nabla }\times \mathbf{B}_{\mathrm{ch}%
}\left( \mathbf{r}|\tilde{\kappa}\right) =\mathbf{j}^{\left( 1\right)
}\left( \mathbf{r}\right) $, we come to the wrong relation $\boldsymbol{%
\nabla }\times \mathbf{B}_{\mathrm{string}}^{\mp }\left( \mathbf{r}|\tilde{%
\kappa}\right) =\mathbf{0}$ in disagreement with Eq. (\ref{sb13}). In short,
the inconsistency is nested in the fact that the string-encircling current (%
\ref{sb13}), whose role is to guarantee the vanishing of the total magnetic
flux, i.e., fulfillment of Bianchi identity, is not present in initial field
equations, neither it is an induced current -- unlike $\mathbf{j}^{\left(
1\right) }\left( \mathbf{r}\right) $.

More comments on this point are in order. Suppose, we would work with field
equations via vector-potentials from the very beginning to intrinsically
guarantee the Bianchi identities. To this end, replace $\mathbf{B}^{\left(
1\right) }\left( \mathbf{r}\right) $ by $\boldsymbol{\nabla }\times \mathbf{A%
}\left( \mathbf{r}\right) $ in Eq.(\ref{nnew7b}). Then, referring to the
gauge condition $\boldsymbol{\nabla }\cdot \mathbf{A}\left( \mathbf{r}%
\right) =0$, it becomes%
\begin{equation}
\boldsymbol{\nabla }\times \lbrack \boldsymbol{\nabla }\times \mathbf{A}%
\left( \mathbf{r}\right) ]=\boldsymbol{\nabla }^{2}\mathbf{A}\left( \mathbf{r%
}\right) =\mathbf{j}^{(1)}\left( \mathbf{r}\right) \,.  \label{nu}
\end{equation}%
This inhomogeneous equation has $\mathbf{A}\left( \mathbf{r}\right) =\mathbf{%
A}^{\left( 1\right) }\left( \mathbf{r}\right) $ Eq. (\ref{stringlessA}) as
its (uncharged) solution, while the\ homogeneous equation $\boldsymbol{%
\nabla }^{2}\mathbf{A}\left( \mathbf{r}\right) =\mathbf{0}$\ does not have
any monopole solution. As for $\mathbf{A}_{\mathrm{ch}}^{\left( \pm \right)
}\left( \mathbf{r}|\tilde{\kappa}\right) $, it does not obey Eq.(\ref{nu}),
but on the contrary 
\begin{equation*}
\boldsymbol{\nabla }\times \lbrack \boldsymbol{\nabla }\times \mathbf{A}_{%
\mathrm{ch}}^{\left( \pm \right) }\left( \mathbf{r}|\tilde{\kappa}\right) ]=%
\boldsymbol{\nabla }^{2}\mathbf{A}_{\mathrm{ch}}^{\left( \pm \right) }\left( 
\mathbf{r}|\tilde{\kappa}\right) =\mathbf{j}^{(1)}\left( \mathbf{r}\right) +%
\mathbf{j}_{\mathrm{string}}^{\mp }\left( \mathbf{r}|\tilde{\kappa}\right)
\,,
\end{equation*}%
keeping in mind the gauge condition $\mathbf{\nabla }\cdot \mathbf{A}_{%
\mathrm{ch}}^{\left( \pm \right) }\left( \mathbf{r}|\tilde{\kappa}\right) =0$
fulfilled everywhere as explained below (\ref{Coulomb}). Thus,
the resulting magnetic solution satisfies the relation%
\begin{equation*}
\boldsymbol{\nabla }\times \left[ \mathbf{B}_{\mathrm{string}}^{\mp }\left( 
\mathbf{r}|\tilde{\kappa}\right) +\mathbf{B}_{\mathrm{ch}}\left( \mathbf{r}|%
\tilde{\kappa}\right) \right] =\mathbf{j}^{(1)}\left( \mathbf{r}\right) +%
\mathbf{j}_{\mathrm{string}}^{\mp }\left( \mathbf{r}|\tilde{\kappa}\right)
\,.
\end{equation*}%
We see again that in the framework of pure vector-potential approach, the\
current $\mathbf{j}_{\mathrm{string}}^{\mp }\left( \mathbf{r}|\tilde{\kappa}%
\right) $ is not inherent in our primary (nonlinear Abelian)
electrodynamics, either. Therefore, in order to support the magnetic
monopole rigged by a Dirac string, whose role is to guarantee the vanishing
of the total magnetic flux, i.e. fulfillment of Bianchi identity, it would
be necessary to introduce additively the solenoidal current into the primary
Lagrangian (\ref{L}) as an additional external source $\mathbf{A}\left( 
\mathbf{r}\right) \cdot \mathbf{j}_{\mathrm{string}}^{\mp }\left( \mathbf{r}|%
\tilde{\kappa}\right) $.

\section{Conclusions\label{Conc}}

Within a general nonlinear local electrodynamics (\ref{L}), we have obtained
static electric\ and\ magnetic fields created by an electric point-like
charge $q$ placed in a background of arbitrarily strong constant electric, $%
\overline{\mathbf{E}}$, and magnetic, $\overline{\mathbf{B}}$, fields,
parallel between themselves, by solving (the second pair of) the Maxwell
equations linearized near the background and treated in the approximation of
small nonlinearity, eqs. (\ref{eq6.1}, \ref{nnew7b}).

The point-like charge $q$ excites an induced charge (\ref{eq10}) and (\ref%
{par}) current densities in the equivalent medium formed by the background
and supplies the Maxwell equations with inhomogeneities. Hence corresponding
solutions -- eq. (\ref{eq9.1}) for electric and eq. (\ref{ansa20}) for
magnetic fields -- are nothing but linear responses of the medium to the
imposed charge $q$. Our formulas for responses are determined by
coefficients that are derivatives of the nonlinear part of the Lagrangian (%
\ref{L}) over field invariants, where the background values of the fields
are meant to be substituted after the derivatives have been calculated.
Electric response (\ref{eq9.1}) is characterized by nonzero total charge (%
\ref{eq11}), which is the screened seeded charge $q$, and by scalar
combinations of external fields. Magnetic response\ eq. (\ref{ansa20}) is
characterized by zero total magnetic charge and by pseudoscalar combinations
of external fields. Both responses are radial, but depend upon the angle $%
\theta $ beetween the radius-vector and the direction of external fields;
they are free of singularities upon a line, but have one in the point $r=0$
where charge $q$ is located. So are the scalar (\ref{eq11b}) \ and vector (%
\ref{stringlessA}) potentials.

There are other solutions -- those to homogeneous counterparts of the
Maxwell equations, eqs. (\ref{eq6.1}, \ref{nnew7b}) -- additive to the
responses. These are determined by arbitrary integration constants. Electric
one, eqs. (\ref{eq18}), (\ref{eq19a}) requires an arbitrary pseudoscalar
factor $\tilde{\varkappa}$, since $\boldsymbol{\Upsilon }\left( \mathbf{r}%
\right) $ is a pseudovector field. It is not radial and it has a angular
singularity at $\theta =0,\pi $ apart from the pole in the origin $r=0$. The
total electric charge is zero, but there is a charge concentrated on
infinitely thin thread stretched along axis $\theta =0,\pi $, it has
opposite signs above and below the origin. We call it di-thread. As usual,
the integration constant -- it is $\tilde{\varkappa}$ in the present context
-- is fixed by the charge with the help of Gauss' theorem. Linear charge
density (\ref{exte}) $\tilde{\varkappa}\boldsymbol{\nabla }\cdot \boldsymbol{%
\Upsilon }\left( \mathbf{r}\right) =2\pi \tilde{\varkappa}\delta \left( 
\mathbf{r}_{\perp }\right) /z=\rho _{\mathrm{di-thr}}\left( \mathbf{r}|%
\tilde{\varkappa}\right) $ is to be considered as external parameter and
included into primary action. The scalar potential (\ref{eq11b}) also bears
singularities at $r=0$ and $\theta =0,\pi $. The pattern of lines of force
and equipotential lines\ is presented in Fig. \ref{Fig2}.

Magnetic solution has two parts: radial and not radial. The radial part (\ref%
{bcharged}) is written \ as a combination of the mentioned magnetic response
(\ref{ansa20}) and a nonzero point-like magnetic monopole with charge $q_{M}$%
\ (\ref{s3.2}), the latter being determined by arbitrary integration
constant, a pseudoscalar $\widetilde{\kappa }$ and the pseudoscalar
combination $\widetilde{g}$\ of constant external fields. (The need of a
pseudoscalar is clear already because the standard magnetic monopole field $%
\thicksim q_{M}\mathbf{r}/r^{3}$ must be a pseudovector, while $\mathbf{r}$
is a polar vector. Hence the experimental search of magnetic monopole is, in
fact, a search of a pseudoscalar in nature. In our context, the latter is
proposed by the scalar product of external electric and magnetic fields $%
\overline{\mathfrak{G}}=-\overline{\mathbf{B}}\cdot \overline{\mathbf{E}}$. %
It is the nonlinearity of electrodynamics that is apt for
combining $\overline{\mathfrak{G}}$ with the deviation field when building a
magnetic monopole\footnote{%
Beyond pure electrodynamics, the cosmological pseudoscalar field,
responsible for the expansion of Universe, may be cosidered as forming the
necessary, almost constant, background. In this respect, Ref. \cite{Altshul}
(and references therein) should be paid attention, where magnetic solutions
are studied under such Lorentz-violating conditions.}). Radial magnetic
solution (\ref{bcharged}) has a singularity in $r=0$. Following the Gauss
theorem, we have in a standard way the magnetic charge density $\boldsymbol{%
\nabla }\cdot \mathbf{B}_{\mathrm{ch}}\left( \mathbf{r}|\tilde{\kappa}%
\right) =q_{M}\delta \left( \mathbf{r}\right) $. This relation violates the
Bianchi identity in $r=0$. As a result, the vector potential (\ref{4.12}) is
singular at $\theta =0$ or at $\theta =\pi $ making Dirac string along any
of the half-axes $\theta =0$ or $\theta =\pi $ . Calculating its circulation
around a half-axis of singularity one finds, as usual, magnetic field in the
string $\mathbf{B}_{\mathrm{string}}^{\mp }\left( \mathbf{r}|\tilde{\kappa}%
\right) =\pm q_{M}\Theta \left( \mp z\right) \delta \left( \mathbf{r}_{\perp
}\right) \mathbf{\hat{z}}$,\ (\ref{Bstring}) and the solenoidal current (\ref%
{sb13})\ flowing around the string $\mathbf{j}_{\mathrm{string}}^{\mp
}\left( \mathbf{r}|\tilde{\kappa}\right) =\boldsymbol{\nabla }\times \mathbf{%
B}_{\mathrm{string}}^{\mp }\left( \mathbf{r}|\tilde{\kappa}\right) =\pm q_{M}%
\left[ \mathbf{\hat{x}}\delta \left( x\right) \delta ^{\prime }\left(
y\right) -\mathbf{\hat{y}}\delta ^{\prime }\left( x\right) \delta \left(
y\right) \right] \Theta \left( \mp z\right) $ to support the magnetic field
inside it.

The nonradial part of the magnetic solution is dual to nonradial part of the
electric solution and is given by the same pseudovector field $\boldsymbol{%
\Upsilon }\left( \mathbf{r}\right) $ (\ref{eq19a}), but this time with a
scalar integration constant $\kappa $ as a factor $\mathbf{B}\left( \mathbf{r%
}|\kappa \right) =\kappa \boldsymbol{\Upsilon }\left( \mathbf{r}\right) $.
This is again field of a thin thread carrying a distributed -- now magnetic
-- charge (\ref{magthread}) of both polarities, the total charge being zero.
The pattern of magnetic lines of force of this magnetic di-thread is the
same as the one presented in Fig.\ref{Fig2} for electric lines of force.

Both, the magnetic charge density, the solenoidal current and thread density
are external agents to be included into primary action.

Lastly, it should be emphasized that the solutions of linearized Maxwell's equations (\ref{MaxEq}) for fields studied
 above within the approximation of lowest order in the magnitude of nonlinearity have the singularity structure
 of $\mathcal{V}/r^2$, where $\mathcal{V}$ is a dimensionless quantity. As the iterative equation (\ref{eq4}) does not include any  dimensional constant in its coefficients, there is no other dimensional quantity except $r$ to form fields. Hence
 this structure will retain in approximation of every order, the summation in (\ref{eq3})  touching only the numerator $\mathcal{V}$.
 
\section*{Acknowledgements}

T. C. Adorno acknowledges the support from the XJTLU Research Development
Funding, award no. RDF-21-02-056, and D. M. Gitman thanks CNPq for permanent
support.

\section*{Data Availability Statement}
Data sharing is not applicable to this article as no new data were created
or analyzed in this study.

\end{document}